\documentclass[prb,floatfix,twocolumn,superscriptaddress,showpacs,amsmath,amssymb]{revtex4}

\usepackage{graphicx,color}% Include figure files
\usepackage{dcolumn}% Align table columns on decimal point
\usepackage{bm}% bold math
\newcommand{\pdag}{{\phantom{\dagger}}}
\newcommand{\be}{\begin{equation}}
\newcommand{\ee}{\end{equation}}
\newcommand{\ba}{\begin{eqnarray}}
\newcommand{\ea}{\end{eqnarray}}
\newcommand{\bea}{\begin{eqnarray*}}
\newcommand{\eea}{\end{eqnarray*}}

\begin{document}

\title{Dynamical behavior across the Mott transition of two bands with different bandwidths} 
\author{Michel Ferrero}
\affiliation{INFM-Democritos, National Simulation Center, and International 
School for Advanced Studies (SISSA), I-34014 Trieste, Italy} 
\author{Federico Becca}
\affiliation{INFM-Democritos, National Simulation Center, and International 
School for Advanced Studies (SISSA), I-34014 Trieste, Italy} 
\author{Michele Fabrizio}
\affiliation{INFM-Democritos, National Simulation Center, and International 
School for Advanced Studies (SISSA), I-34014 Trieste, Italy} 
\affiliation{International Centre for Theoretical Physics
(ICTP), P.O.Box 586, I-34014 Trieste, Italy}
\author{Massimo Capone}
\affiliation{INFM-SMC and Dipartimento di Fisica, Universit\`a di Roma ``La Sapienza'', 
Piazzale Aldo Moro 2, I-00185 Roma, Italy}
\affiliation{Istituto dei Sistemi Complessi del CNR, via dei Taurini 19, I-00185 
Roma, Italy}

\date{\today}
\begin{abstract}
We investigate the role of the bandwidth difference in the Mott metal-insulator transition 
of a two-band Hubbard model in the limit of infinite dimensions, by means of a 
Gutzwiller variational wave function as well as by dynamical mean-field theory. 
The variational calculation predicts a two-stage quenching of the charge 
degrees of freedom, in which the narrower band undergoes a Mott transition 
before the wider one, 
both in the presence and in the absence of a Hund's exchange coupling. 
However, this scenario is not fully confirmed by the dynamical mean-field theory calculation, 
which shows that, although the quasiparticle residue of the narrower band is zero  
within our numerical accuracy, low-energy spectral weight still exists 
inside the Mott-Hubbard gap, concentrated
into two peaks symmetric around the chemical potential. This spectral weight 
vanishes only when the wider band ceases to conduct too. 
Although our results are compatible with several scenarios, e.g., a narrow gap 
semiconductor or a semimetal, we argue that the most plausible one 
is that the two peaks coexist with a narrow resonance tied at the chemical potential,
with a spectral weight below our numerical accuracy. 
This quasiparticle resonance is expected to vanish when the wider band undergoes the Mott 
transition.
\end{abstract}
\maketitle

\section{Introduction}\label{sc:intro}

Unlike in single-band models, the Mott metal-insulator transition
(MIT) in multi-orbital strongly correlated systems generically
involves other energy scales besides the short range Coulomb repulsion
$U$ and the {\it bare} electron bandwidth. They include, for instance,
the Coulomb exchange $J$ which produces the Hund's rules, any
crystal field or Jahn-Teller effect splitting the orbital degeneracy,
and possibly bandwidth differences between the orbitals.  There are
many theoretical works making use of the so-called Dynamical Mean-Field
Theory (DMFT)~\cite{georges1996} which analyze the role of the exchange
$J$,~\cite{Bulla} the crystal field splitting,~\cite{Manini} and the
Jahn-Teller effect.~\cite{Han,Massimo} All these analysis suggest that these
perturbations, which have the common feature of splitting multiplets
at fixed charge, are amplified near the MIT, leading for instance to
an appreciable shift of the transition towards lower
$U$'s~\cite{Bulla} or to the appearance of anomalous phases just
before the MIT.~\cite{Massimo} This behavior is not surprising,
since the more the electronic motion is slowed down, i.e., the
longer is the time electrons stay localized around a site, the larger
is the chance to get advantage of multiplet-splitting mechanisms.

On the contrary, the role of different bandwidths for nearly
degenerate orbitals is less predictable, since the Coulomb charge
repulsion only cares about the total number of electrons at a given
site, while it is not concerned with the orbital they sit in.
Recently, this issue has been addressed in a two-band Hubbard model by
DMFT, yet leading to controversial results. Liebsch has suggested, on 
the basis of a DMFT calculation using quantum Monte Carlo as impurity solver, 
that both with
$J=0$ and $J\not =0$ the two orbitals undergo a common MIT, whatever
the difference between their bandwidths.~\cite{liebsch2003,liebsch2004} 
This conclusion has been questioned by Koga and coworkers,~\cite{koga2004} 
again by DMFT, using, however, exact
diagonalization instead of quantum Monte Carlo. Their conclusion is that the MIT is
unique only if $J=0$. On the contrary, when $J\not = 0$, there is a
first transition at which the orbital with smaller bandwidth becomes
insulating, followed at larger values of the interaction by a
second transition at which the other orbital ceases to conduct as well.
This two-stage quenching of the charge degrees of freedom has been
named Orbital-Selective Mott Transition (OSMT) by those authors.
Although the coexistence of localized $f$-electrons and itinerant
$d$-electrons is not unusual in rare-earth compounds, the conclusions
of Ref.~\onlinecite{koga2004} are a bit surprising in the case of
degenerate orbitals. Indeed, the Coulomb exchange-splitting, rather
than favoring an OSMT, should na\"{\i}vely oppose to it, since $J$
competes against the angular momentum quenching due to the different
bandwidths.

In this work, we attempt to clarify this issue by means of a variational analysis based 
on Gutzwiller wave functions, by standard DMFT calculations as well as by 
an approximate DMFT projective technique. The paper is organized as follows. 
In Section~\ref{sc:The model}, we introduce the two-band model and discuss general properties.  
In Section~\ref{sc:gutzwiller}, we apply a variational technique based on Gutzwiller-type
trial wave functions to analyze the ground state of the Hamiltonian.  
A full DMFT analysis is presented in Section~\ref{sc:dmft}. As a guide to interpret the 
DMFT spectral functions, in Section~\ref{sc:AIM}, we show the density of state 
obtained by Wilson Numerical Renormalization Group~\cite{NRG} 
of the Anderson impurity model onto which the lattice model maps within DMFT.
In Section~\ref{sc:psct}, we present an approximate DMFT solution  
obtained by projecting out self-consistently high-energy degrees of freedom, which allow 
a better low-energy description. Conclusions are drawn in 
Section~\ref{sc:conclusion}.

\section{The model}\label{sc:The model}      

We consider a two-band Hubbard model at half-filling described by the
Hamiltonian
\begin{eqnarray}\label{eq:hamilt}
  \mathcal{H} &=&    - \sum_{\langle i,j \rangle, \sigma}\,\sum_{a=1}^2\, t_a \,  
    f^\dagger_{i, a \sigma} f^{\phantom{\dagger}}_{j, a \sigma} + H.c. \nonumber \\
&&    + \frac{U}{2} \sum_{i} \, \left(n_i - 2\right)^2 
+ \mathcal{H}_{\rm exch},
\end{eqnarray}
where $f^\dagger_{i, a \sigma}$ creates an electron at site $i$ in
orbital $a=1,2$ with spin $\sigma$, $n_{i\, a} =
\sum_{\sigma}\,f^\dagger_{i, a \sigma} f^{\phantom{\dagger}}_{i, a
  \sigma}$ is the occupation number at site $i$ in orbital $a$, and
$n_i = n_{i\, 1} + n_{i\, 2}$ is the total occupation number. The
explicit expression of the Coulomb exchange $\mathcal{H}_{\rm exch}$ is
\ba
\mathcal{H}_{\rm exch} &=& \frac{J}{2} \sum_i\, \left(n_{i\, 1} - n_{i\, 2}\right)^2 
\nonumber\\
&& + \frac{J}{2}\sum_i\, \sum_{\sigma, \sigma'} \, 
f^\dagger_{i,1\sigma}f^\dagger_{i,1\sigma'}f^\pdag_{i,2\sigma'}f^\pdag_{i,2\sigma}
+ H.c. \label{Hexch-pair hopping}\\
&& + \frac{J}{4}\sum_{\sigma, \sigma'} 
f^\dagger_{i,1\sigma}f^\pdag_{i,2\sigma}f^\dagger_{i,2\sigma'}f^\pdag_{i,1\sigma'}
+ H.c.\nonumber \\
&\equiv& 2J\,\sum_i\, \left(T_{i\, x}^2 + T_{i\, z}^2\right),\label{Hexch-spin exchange}
\ea
where
\be
T_{i\,\alpha} = \frac{1}{2}\sum_{a, b}\,\sum_\sigma \, 
f^\dagger_{i,a\sigma}\, \tau^\alpha_{ab}\, 
f^\pdag_{i,b\sigma}
\label{pseudo spin}
\ee
are pseudo-spin-1/2 operators, with $\tau^\alpha$ the Pauli matrices,
$\alpha=x,y,z$.  In what follows, we always take $0<t_2\leq t_1$.  Let
us start by discussing some general properties of this Hamiltonian.
 
If $U\gg t_1$, the model describes a Mott insulator in which two
electrons localize on each site. For $J>0$, the atomic two-electron ground state
is the spin triplet, followed at energy $2J$ by the two degenerate
singlets (we drop the site index):
\[
\sqrt{\frac{1}{2}}\left(
f^\dagger_{1\uparrow}f^\dagger_{2\downarrow} 
+ f^\dagger_{2\uparrow}f^\dagger_{1\downarrow}\right)\,|0\rangle,\;\; 
\sqrt{\frac{1}{2}}\left(
f^\dagger_{1\uparrow}f^\dagger_{1\downarrow} 
- f^\dagger_{2\uparrow}f^\dagger_{2\downarrow}\right)\,|0\rangle,
\]
and finally at energy $4J$ by the singlet:
\[
\sqrt{\frac{1}{2}}\left(
f^\dagger_{1\uparrow}f^\dagger_{1\downarrow} 
+ f^\dagger_{2\uparrow}f^\dagger_{2\downarrow}\right)\,|0\rangle. 
\]
Hence, the Mott insulator for very large $U$, specifically $t_1^2/U\ll
J$, is effectively a spin-1 Heisenberg model where, at any site, each
orbital is occupied by one electron, the two electrons being bound
into a spin-triplet configuration.  Within the OSMT scenario, below
some critical repulsion, defined in the following as $U_1$, electrons
in orbital 1 start moving, while one electron per site remains
localized in orbital 2.  Only below a lower $U_2<U_1$, electrons in
orbital 2 delocalize too.  In this particular example with a
half-filled shell, the Coulomb exchange does not conflict with the
OSMT, since $\mathcal{H}_{\rm exch}$ favors single occupancy of each
orbital.  Yet, one may wonder about the role of the pair-hopping
term~(\ref{Hexch-pair hopping}) which can transfer electrons from the
delocalized orbital to the localized one.

\section{Gutzwiller variational technique}\label{sc:gutzwiller}

Let us start by a variational analysis of the ground state of the
Hamiltonian~(\ref{eq:hamilt}). In particular, we are going to use the
Gutzwiller variational approach, which is one of the simplest way to
include electronic correlations into a many-body wave function. In
what follows, we briefly introduce this variational technique for a
generic multi-orbital model, and later we apply it to our specific
example.

Let us consider in general a $k$-orbital Hamiltonian that contains, besides the
hopping term
\be
\mathcal{H}_0 = - \sum_{\langle i,j \rangle, \sigma}\,\sum_{a=1}^k\, t_a \,  
    f^\dagger_{i, a \sigma} f^{\phantom{\dagger}}_{j, a \sigma} + H.c.,
\label{Gutz:Hhop}
\ee
an on-site interaction of the general form 
\be
\mathcal{H}_{\rm int} = \sum_i\, \sum_{n,\Gamma} \, U(n,\Gamma) \, 
\mathcal{P}_i(n,\Gamma),
\label{Gutz:Hint}
\ee
where $\mathcal{P}_{i}(n,\Gamma)=|i;n,\Gamma\rangle\langle
i;n,\Gamma|$ is the projector operator onto the site-$i$ configuration
$\Gamma$ with $n$ electrons.  If $|\Psi_0\rangle$ is the Fermi-sea
Slater determinant of the non-interacting Hamiltonian, the Gutzwiller
wave function $|\Psi\rangle$ is defined through
\be
|\Psi\rangle = \mathcal{P}_G\, |\Psi_0\rangle = 
\prod_i \mathcal{P}_{i\, G}\,|\Psi_0\rangle,
\label{def:wavefunction}
\ee
where the operator $\mathcal{P}_{i\, G}$ acts on site $i$ and is given
by
\be
\mathcal{P}_{i\, G}  = \sum_{n,\Gamma} \, 
\lambda_{n\Gamma}\,\mathcal{P}_{i}(n,\Gamma).
\label{def:PG}
\ee
The $\lambda_{n\Gamma}$'s in~(\ref{def:PG}) are variational parameters
which modify the weights of the on-site configurations
$|i;n,\Gamma\rangle$ in accordance with the interaction~(\ref{Gutz:Hint}).
Following Ref.~\onlinecite{Gebhard}, we assume,
without loss of generality,~\cite{Attaccalite} that
\ba
\langle \Psi_0| \mathcal{P}_{i\, G}^2 |\Psi_0\rangle &=& 
\langle \Psi_0| \Psi_0\rangle = 1,\label{norma}\\
\langle \Psi_0| \, \mathcal{P}_{i\, G}^2\, 
f^\dagger_{i,a \sigma} f^\pdag_{i,b \sigma'}\, 
|\Psi_0\rangle &=& 
\langle \Psi_0|  
f^\dagger_{i,a\sigma} f^\pdag_{i,b \sigma'}   |\Psi_0\rangle\nonumber \\ 
&=& \frac{N}{2k}\,\delta_{ab}\,\delta_{\sigma\sigma'},\label{density-matrix}
\ea
where $N$ is the average occupation per site within the Fermi sea, in
which all orbitals are equally occupied by our choice of the hopping
term~(\ref{Gutz:Hhop}).  These two conditions allow an analytic
evaluation of any average value over the correlated wave function in
infinite dimensions.~\cite{Gebhard} In particular,
one can show that~(\ref{norma}) is satisfied by
\be
\lambda_{n\Gamma}^2 = \frac{P(n,\Gamma)}{P^{(0)}(n,\Gamma)},
\label{norma2}
\ee
where
\be
P(n,\Gamma) = \langle \Psi |{\cal{P}}_i(n,\Gamma)|\Psi\rangle,
\ee
is the correlated probability distribution of the on-site
configuration $\Gamma$ with $n$ electrons and
\be
P^{(0)}(n,\Gamma) = \langle \Psi_0 |{\cal{P}}_i(n,\Gamma)|\Psi_0\rangle,
\ee
is the uncorrelated one. Therefore, rather than using the variational
parameters $\lambda_{n\Gamma}$'s, one can directly work with the
correlated probability distribution $P(n,\Gamma)$ which satisfies
\[
\sum_{n,\Gamma}\, P(n,\Gamma) = 1,\;\;
\sum_{n,\Gamma}\, n\, P(n,\Gamma) = N.
\]
By the above definitions, one can demonstrate~\cite{Gebhard} that the
average value of the hopping in infinite dimensions reduces to
\ba
\langle \Psi | f^\dagger_{i,a\sigma} f^\pdag_{j,a\sigma} |\Psi\rangle 
&=& \langle \Psi_0 | 
\mathcal{P}_{i\, G}\,f^\dagger_{i,a\sigma}\, \mathcal{P}_{i\, G} \,
\mathcal{P}_{j\, G}\,f^\pdag_{j,a\sigma}\,\mathcal{P}_{j\, G} |\Psi_0\rangle \nonumber \\
&=& Z_a \,\langle \Psi_0 | f^\dagger_{i,a\sigma} f^\pdag_{j,a\sigma} 
|\Psi_0\rangle,
\ea
where the reduction factor $Z_a$ can be evaluated through 
\be
\sqrt{Z_a} = \frac{2k}{N} \langle \Psi_0 | 
\mathcal{P}_{i\, G}\,f^\dagger_{i,\alpha\sigma}\, \mathcal{P}_{i\, G} \,
f^\pdag_{i,\alpha\sigma}|\Psi_0\rangle.
\ee
In addition, the average value of the on-site interaction is found to
be
\be
\langle \Psi | \mathcal{H}_{\rm int} |\Psi\rangle = 
\sum_i\,\sum_{n,\Gamma}\, U(n,\Gamma) \, P(n,\Gamma).
\ee 

\subsection{Results for $J=0$} 
Let us apply this technique to our model~(\ref{eq:hamilt}), starting
with the simpler case where $J=0$.  We define as $P(0)$ and $P(4)$ the
correlated probabilities of a site occupied by zero and four
electrons, respectively. $P(1,+)$ and $P(1,-)$ are instead the
probabilities of a singly-occupied site with one electron in orbital 2
and 1, respectively.  Analogously, $P(3,+)$ and $P(3,-)$ are the
probabilities of a site occupied by three electrons, with one hole in
orbital 1 or 2, respectively. Finally, concerning doubly occupied
sites, $P(2,1)$ is the probability of the spin-triplet configuration,
$P(2,0)$ of the inter-orbital spin-singlet, $P(2,+)$ and $P(2,-)$ of
the spin singlets with two electrons in orbital 2 and 1, respectively.
By particle-hole symmetry, we have that
\bea
P(0) &=& P(4),\\
P(1,+) &=& P(3,-),\\
P(1,-) &=& P(3,+),\\
P(2,+) &=& P(2,-).
\eea
After some simple, but lengthy algebra, we find that the hopping energy
reduction factors are
\bea
\sqrt{Z_1} &=& 2\,\sqrt{2}\, \sqrt{P(0)P(1,-)} + 
\sqrt{2}\,\sqrt{P(1,-)P(2,-)}\\
&& + 
\sqrt{2}\,\sqrt{P(1,-)P(2,+)} + \sqrt{2}\,\sqrt{P(1,+)P(2,0)}\\
&& + \sqrt{6}\,\sqrt{P(1,+)P(2,1)},\\
\sqrt{Z_2} &=& 2\,\sqrt{2}\, \sqrt{P(0)P(1,+)} + 
\sqrt{2}\,\sqrt{P(1,+)P(2,+)}\\
&& + 
\sqrt{2}\,\sqrt{P(1,+)P(2,-)} + \sqrt{2}\,\sqrt{P(1,-)P(2,0)}\\
&& + \sqrt{6}\,\sqrt{P(1,-)P(2,1)}. 
\eea
Since $J=0$, the two-electron configurations with one electron in
each orbital are equally probable, namely $P(2,1)= 3P(2,0)$. Therefore,
one can use the following parametrization
\bea 
P(1,+) &=& P(1)\, \cos^2\phi,\\
P(1,-) &=& P(1)\, \sin^2\phi,\\
P(2,1) &=& P(2)\, \frac{3}{4}\,\cos^2\theta,\\
P(2,0) &=& P(2)\, \frac{1}{4}\,\cos^2\theta,\\
P(2,+) &=& P(2,-) = P(2)\, \frac{1}{2}\,\sin^2\theta.
\eea
The normalization condition now reads
\be
2\,P(0) + 2\,P(1) + P(2) = 1,
\label{norm3}
\ee
and the $Z$-reduction factors can be written as 
\bea
\sqrt{Z_1} &=& 2\,\sqrt{2}\,\sin\phi\,\sqrt{P(0)P(1)} \\
&& +
\sqrt{P(1)P(2)}\,\left[
2\,\sin\phi\,\sin\theta + \sqrt{8}\, \cos\phi\,\cos\theta\right],\\
\sqrt{Z_2} &=& 2\,\sqrt{2}\,\cos\phi\,\sqrt{P(0)P(1)} \\
&& + 
\sqrt{P(1)P(2)}\,\left[
2\,\cos\phi\,\sin\theta + \sqrt{8}\, \sin\phi\,\cos\theta\right]. 
\eea
If we define
\bea
T &=& 
\frac{1}{V}\langle \Psi_0 | \sum_{\langle i,j \rangle, \sigma}\, f^\dagger_{i,1\sigma} 
f^\pdag_{j,1\sigma} + H.c. |\Psi_0\rangle \\
&=&  \frac{1}{V}\langle \Psi_0 | \sum_{\langle i,j \rangle, \sigma}\, f^\dagger_{i,2\sigma} 
f^\pdag_{j,2\sigma} + H.c. |\Psi_0\rangle,
\eea
the average value per site of the hopping operator in the Fermi sea,
then the variational energy of the Gutzwiller wave function in
infinite dimensions is
\be
E = -t_1\, T\, Z_1 -t_2\,T\,Z_2 + U\,P(1) + 4\,U\,P(0).
\label{E-var}
\ee
Here, the $Z_i$'s are functionals of the probability distribution
$P(n)$, $n=0,1,2$, with the normalization~(\ref{norm3}), as well as of
the two angles $\phi,\theta\in [0,\pi/2]$.
 
One can proceed analytically a bit further. It is known that near a
Mott transition and within the Gutzwiller wave function approach, one
can safely neglect $P(0)=P(4)$.~\cite{Attaccalite} Within this
approximation, $P(2)=1-2\,P(1)$ and
\bea
\sqrt{Z_1} &=& \sqrt{P(1)P(2)}\,\left[
2\,\sin\phi\,\sin\theta + \sqrt{8}\, \cos\phi\,\cos\theta\right],\\
\sqrt{Z_2} &=& \sqrt{P(1)P(2)}\,\left[
2\,\cos\phi\,\sin\theta + \sqrt{8}\, \sin\phi\,\cos\theta\right].
\eea
We denote $P(1)=d/2$, hence $P(2)=1-d$, so that the variational
energy~(\ref{E-var}) becomes
\[
E = - T\, \frac{d\,(1-d)}{2}\, f(\phi,\theta) + \frac{U}{2}\, d
\]
where 
\bea
f(\phi,\theta) &=& t_1 \,  \left[
2\,\sin\phi\,\sin\theta + \sqrt{8}\, \cos\phi\,\cos\theta\right]^2 \\
&& + t_2\, \left[
2\,\cos\phi\,\sin\theta + \sqrt{8}\, \sin\phi\,\cos\theta\right]^2.
\eea
The optimal value for $d$ is 
\[
d_* = \frac{T\,f(\phi,\theta) - U}{2\,T\,f(\phi,\theta)},
\]
and the variational energy becomes:
\[
E[\phi,\theta] = - \frac{\left[T\,f(\phi,\theta) - U\right]^2}{8\, T\, f(\phi,\theta)}.
\]
At given $\phi$ and $\theta$, the Mott transition   
at which both orbitals localize occurs when $d_*=0$,
namely when
\[
U_c(\phi,\theta) = T\,f(\phi,\theta).
\]
The most stable solution is the one which maximizes $f$. An OSMT
corresponds to a situation in which the Mott transition occurs with  
orbital 2 being already strictly singly-occupied, namely with $\phi=\theta=0$.  
This solution is an extremum of $f$. Yet, one has to check 
whether it is also a maximum. We find
that this is indeed the case whenever
\be
t_2 \leq \frac{1}{5}\, t_1.
\label{Gutz:critical ratio}
\ee
Therefore, within the Gutzwiller variational technique, an OSMT can
occur even in the absence of Coulomb exchange,
provided~(\ref{Gutz:critical ratio}) is satisfied.

\begin{figure}
\includegraphics[width=0.48\textwidth]{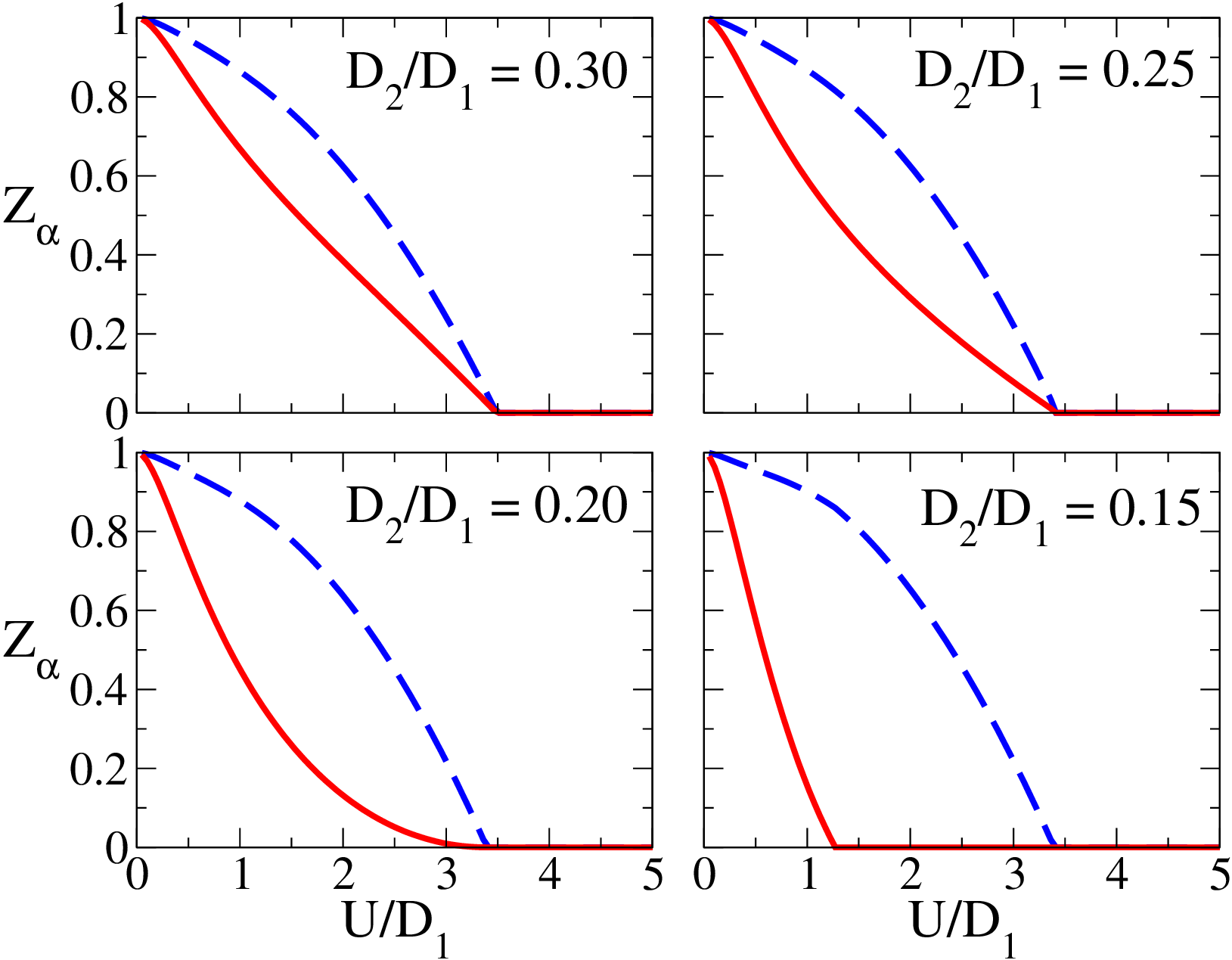}
\caption{
  Reduction factors $Z_a$ for the wide band (dashed line) and for the
  narrow band (continuous line) obtained with the Gutzwiller wave
  function for different ratios of the bandwidth $D_2 / D_1$ and $J =
  0$.}
\label{fg:ZGutzJ000}
\end{figure}

We optimized~(\ref{E-var}) numerically, considering an infinite
coordination Bethe lattice at half-filling. The free density of state
is given by
\begin{equation}\label{bethedensity}
  \rho_a(\epsilon) =
    \frac{\sqrt{4 t^2_a - \epsilon^2}}{2 \pi t^2_a},
\end{equation}
and, in this case, $T = 8/(3 \pi)$. The half-bandwidth of each band is
$D_a = 2 t_a$, and our unit of energy is $D_1$. The reduction factors
$Z_1$ and $Z_2$ are shown in Fig.~\ref{fg:ZGutzJ000}, for different
ratios $D_2 / D_1$. The results indeed confirm the analytical
calculation~(\ref{Gutz:critical ratio}), 
displaying two distinct transitions when $D_2 / D_1 < 0.20$.

%%%%%%%%%%%%%%%%%%%%%%%%%%%%%%%%%%%%%%%%%%%%%%%%%%%%%%%%%%%%%%%%%%%%%%%%%%%
%%%%%%%%%%%%%%%%%%%%%%%%%%%%%%%%%%%%%%%%%%%%%%%%%%%%%%%%%%%%%%%%%%%%%%%%%%%

\subsection{Results for $J\not =0$}
\begin{figure}
\includegraphics[width=0.48\textwidth]{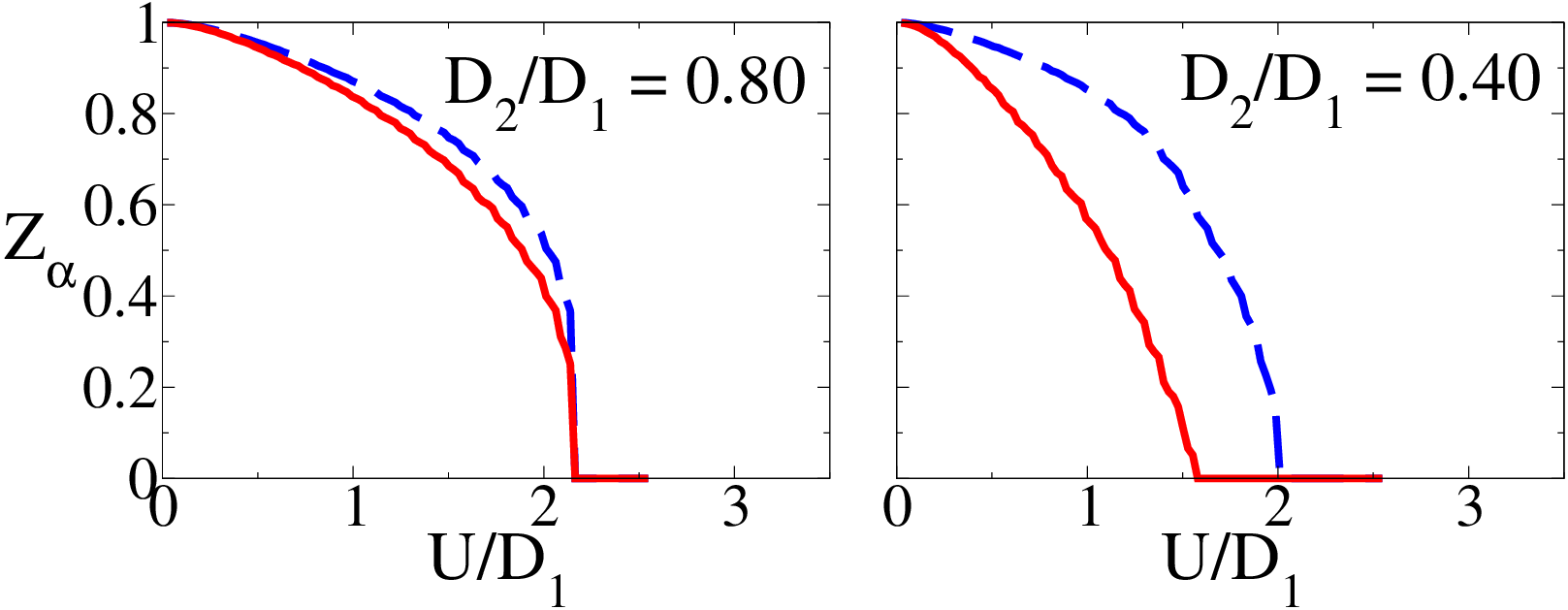}
\caption{
  Reduction factors $Z_a$ for the wide band (dashed line) and the
  narrow band (continuous line) obtained with the Gutzwiller wave
  function for different ratios of the bandwidth $D_2 / D_1$ and $J /
  U = 0.10$.}
\label{fg:ZGutzJ010}
\end{figure}

\begin{figure}
\includegraphics[width=0.48\textwidth]{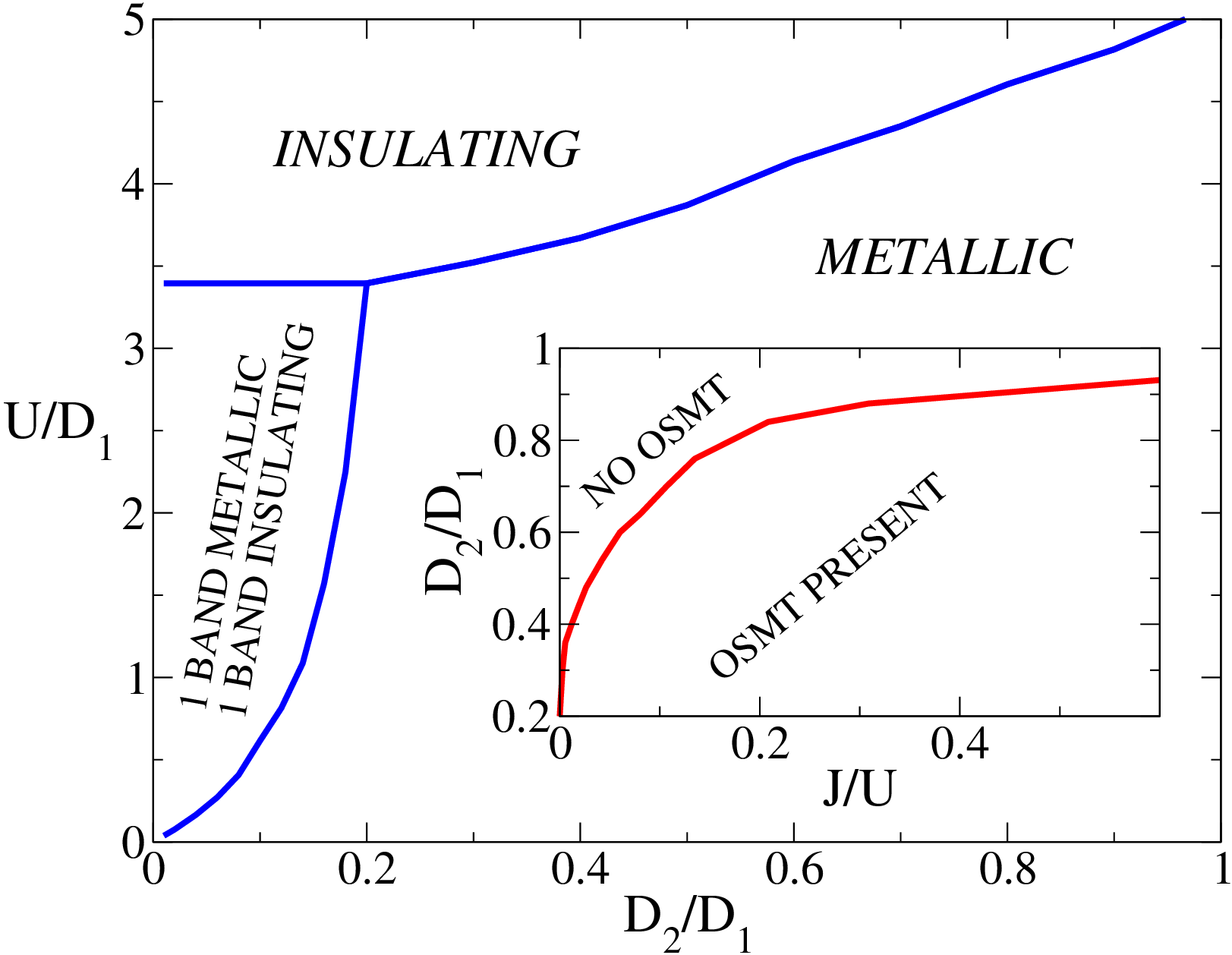}
\caption{
  Phase diagram obtained within the Gutzwiller approximation in the
  absence of the Hund's coupling. Inset: the critical ratio $D_2 /
  D_1$ below which an OSMT is observed as a function of $J / U$.}
\label{fg:phasediag}
\end{figure}

Let us now move to the more complicated case of $J\not = 0$.  The
Hund's rule coupling acts only within the two-electron configurations
and can be written as
\ba
\mathcal{H}_J &=& \sum_i\, -\frac{4}{3} J\, |i;2,1\rangle\langle i;2,1| 
+ \frac{2}{3}J\, |i;2,0\rangle\langle i;2,0| \nonumber \\
&& + \frac{5}{3} J\,\left(\,
|i;2,+\rangle\langle i;2,+| + 
|i;2,-\rangle\langle i;2,-| \, \right) \nonumber \\
&& + J \,\left(\,
|i;2,+\rangle\langle i;2,-| + 
|i;2,-\rangle\langle i;2,+| \, \right),
\label{Gutzw-Hund}
\ea
using the previous notation for the two-electron configurations.
Since it is not diagonal in the representation which we have used so
far, we are forced to generalize the Gutzwiller correlator~(\ref{def:PG})
into~\cite{Attaccalite}
\be
\mathcal{P}_{i \, G} \rightarrow 
\mathcal{P}_{i \, G} + \lambda_{2\,\perp} 
\left(|2,+\rangle\langle 2,-| + |2,-\rangle\langle 2,+|\right).
\label{def:PG2}
\ee
Consequently,
\bea
P(2,+) &=& \lambda_{2\,+}^2\, P^{(0)}(2,+) + \lambda_{2\,\perp}^2\, P^{(0)}(2,-),\\
P(2,-) &=& \lambda_{2\,-}^2\, P^{(0)}(2,-) + \lambda_{2\,\perp}^2\, P^{(0)}(2,+).
\eea
Since by particle-hole symmetry $P(2,+)=P(2,-)$, as well as
$P^{(0)}(2,+)=P^{(0)}(2,-)$, then $\lambda_{2\,+}=\lambda_{2\,-}\equiv
\lambda_{2\,\pm}$ and hence
\[
\lambda_{2,\pm}^2 + \lambda_{2,\perp}^2 = \frac{P(2,+)}{P^{(0)}(2,+)}.
\]
In addition, the following matrix element is non zero
\be
A(2,\pm) \equiv \langle \Psi|\, 2,+\rangle\langle 2,-\, |\Psi\rangle = 
2\,\lambda_{2,\perp}\, \lambda_{2,\pm}\,P^{(0)}(2,+), 
\ee
so that 
\bea
\lambda_{2,\pm} &=& \frac{1}{2\sqrt{P^{(0)}(2,+)}}\left[ 
\sqrt{P(2,+)+A(2,\pm)} \right. \\
&& \left. + \sqrt{P(2,+)-A(2,\pm)}\right],\\
\lambda_{2,\perp} &=& \frac{1}{2\sqrt{P^{(0)}(2,+)}}\left[ 
\sqrt{P(2,+)+A(2,\pm)} \right. \\
&& \left. - \sqrt{P(2,+)-A(2,\pm)}\right].
\eea
We notice that $|A(2,\pm)|\leq P(2,+)$.  The hopping reduction factors
are consequently modified into
\bea
\sqrt{Z_1} &=& 2\sqrt{2}\,\sqrt{P(0)P(1,-)} \\
&& + \sqrt{2}\,\sqrt{P(1,-)}\left[ \sqrt{P(2,+)+A(2,\pm)}\right. \\ 
&& \left. + \sqrt{P(2,+)-A(2,\pm)}\right]\\
&& + \sqrt{2}\,\sqrt{P(1,+)P(2,0)} + 
\sqrt{6}\,\sqrt{P(1,+)P(2,1)},\\
\sqrt{Z_2} &=& 2\sqrt{2}\,\sqrt{P(0)P(1,+)} \\
&& + \sqrt{2}\,\sqrt{P(1,+)}\left[ \sqrt{P(2,+)+A(2,\pm)}\right. \\ 
&&\left. + \sqrt{P(2,+)-A(2,\pm)}\right]\\
&& + \sqrt{2}\,\sqrt{P(1,-)P(2,0)} + 
\sqrt{6}\,\sqrt{P(1,-)P(2,1)}. 
\eea
The average value of the Hund's coupling is 
\ba
E_J &=& -\frac{4}{3}\,J\,P(2,1) + \frac{8}{3}\,J\,\left[P(2,+)+A(2,\pm)\right]
\nonumber\\
&& + \frac{2}{3}\,J\,\left[P(2,0)+P(2,+)-A(2,\pm)\right],
\label{EJ}
\ea
and that of the Hubbard repulsion
\be
E_U = U\,\left[P(1,+)+P(1,-)\right] + 4\,U\,P(0), 
\label{EU}
\ee
thus, the variational energy to be minimized is 
\be
E = -t_1\,T\,Z_1 -t_2\,T\,Z_2 +E_J +E_U.
\label{E-var2}
\ee
By the numerical minimization of the variational energy, we find that
the critical ratio $D_2/D_1$ for an OSMT increases when $J\not = 0$
from the value 0.2 found for $J=0$. In Fig.~\ref{fg:ZGutzJ010}, we show
$Z_1$ and $Z_2$ as obtained by numerical minimization of~(\ref{E-var2}),
for two ratios of $D_2 / D_1$ and $J/U = 0.10$.
Additional calculations allowed to draw the phase diagram within the
Gutzwiller variational approach, see Fig.~\ref{fg:phasediag}. As is
apparent from the inset, the introduction of the Hund's coupling
increases the value of the critical ratio $D_2 / D_1$.
We also notice that the Mott transition at which both bands localize becomes 
first order in the presence of $J$, which, however, might be a pathology of 
the Gutzwiller wave function.~\cite{Attaccalite}

%%%%%%%%%%%%%%%%%%%%%%%%%%%%%%%%%%%%%%%%%%%%%%%%%%%%%%%%%%%%%%%%%%%%%%%%%%%
%%%%%%%%%%%%%%%%%%%%%%%%%%%%%%%%%%%%%%%%%%%%%%%%%%%%%%%%%%%%%%%%%%%%%%%%%%%

%%%%%%%%%%%%%%%%%%%%%%%%%%%%%%%%%%%%%%%%%%%%%%%%%%%%%%%%%%%%%%%%%%%%%%%%%%%
%%%%%%%%%%%%%%%%%%%%%%%%%%%%%%%%%%%%%%%%%%%%%%%%%%%%%%%%%%%%%%%%%%%%%%%%%%%

In conclusion, we find that the Gutzwiller variational technique
predicts an OSMT both for $J=0$ and $J\not = 0$, provided $D_1/D_2$ is
smaller than a critical value which increases with $J$. 

\section{Dynamical Mean-Field Theory}\label{sc:dmft}

To have further insights on the quality of the Gutzwiller wave function
in infinite dimensions, we have performed an extensive DMFT
calculation for the same Hamiltonian. DMFT is a
non-perturbative approach that neglects spatial correlations but
retains fully time correlations and becomes exact for
infinite-coordination lattices. In this limit, the lattice 
problem~(\ref{eq:hamilt}) can be mapped onto an Anderson impurity model
supplemented by a self-consistency condition. For simplicity, we
consider an infinite coordination Bethe lattice, as in the Gutzwiller
variational approach, with a bare density of states given
by~(\ref{bethedensity}). Again, $D_a = 2 t_a$ is the half-bandwidth of
band $a$. The Anderson impurity model onto which the lattice model
maps within DMFT is
\begin{eqnarray}\label{eq:Hand}
  \mathcal{H}_\mathrm{AM} &=&
    \sum_{k, a, \sigma}
    \epsilon_{k a}
    c^\dagger_{k a \sigma} c^{\phantom{\dagger}}_{k a \sigma}
    + \sum_{k, a, \sigma} V_{k a}
    \left(f^\dagger_{a \sigma} c^{\phantom{\dagger}}_{k a \sigma}
    + H.c.\right) \nonumber \\
    &+& \frac{U}{2} (n_f - 2)^2
    + 2J \left(T_x^2 + T_z^2\right),
\end{eqnarray}
where $T_\alpha$, $\alpha=x,y,z$, are the pseudo-spin operators~(\ref{pseudo spin})
for the impurity.  The self-consistency condition
relates the impurity Green's function for orbital $a$, $G_a$, to the
parameters $\epsilon_{k a}$ and $V_{k a}$ through
\begin{equation}\label{eq:selfcons}
  t_a^2 G_a(i \omega_n) = \sum_{ k }
    \frac{V_{k a}^2}
    {i \omega_n - \epsilon_{k a}}.
\end{equation}

We solve the Anderson impurity model using an exact diagonalization
method~\cite{caffarel1994} at zero temperature. The continuous
conduction-electron bath is modeled by a finite number of parameters
$\epsilon_{k a}$ and $V_{k a}$ ($k = 1, \ldots, n_s-1$). In our
calculations, we considered $n_s = 6$ and $4$ (not shown). 
The self-consistency~(\ref{eq:selfcons}) is implemented through a 
fitting procedure along the imaginary axis. 
To this end, we discretize the axis into Matsubara
frequencies $i \omega_n = (2n + 1) \pi / \beta$, where $\beta$ is a
fictitious temperature that we have set to $\beta = 500$. 
Moreover, the smallest frequency $\omega_{min}$ has been determined by
the smallest pole in the continued fraction expansion of the Green's 
function.~\cite{moeller1995} Note that this procedure sometimes leads 
to different minimum cutoff frequencies for the two orbitals.
In the following, we will work in units of $D_1$.

%%%%%%%%%%%%%%%%%%%%%%%%%%%%%%%%%%%%%%%%%%%%%%%%%%%%%%%%%%%%%%%%%%%%%%%%%%%
\begin{figure}
\includegraphics[width=0.48\textwidth]{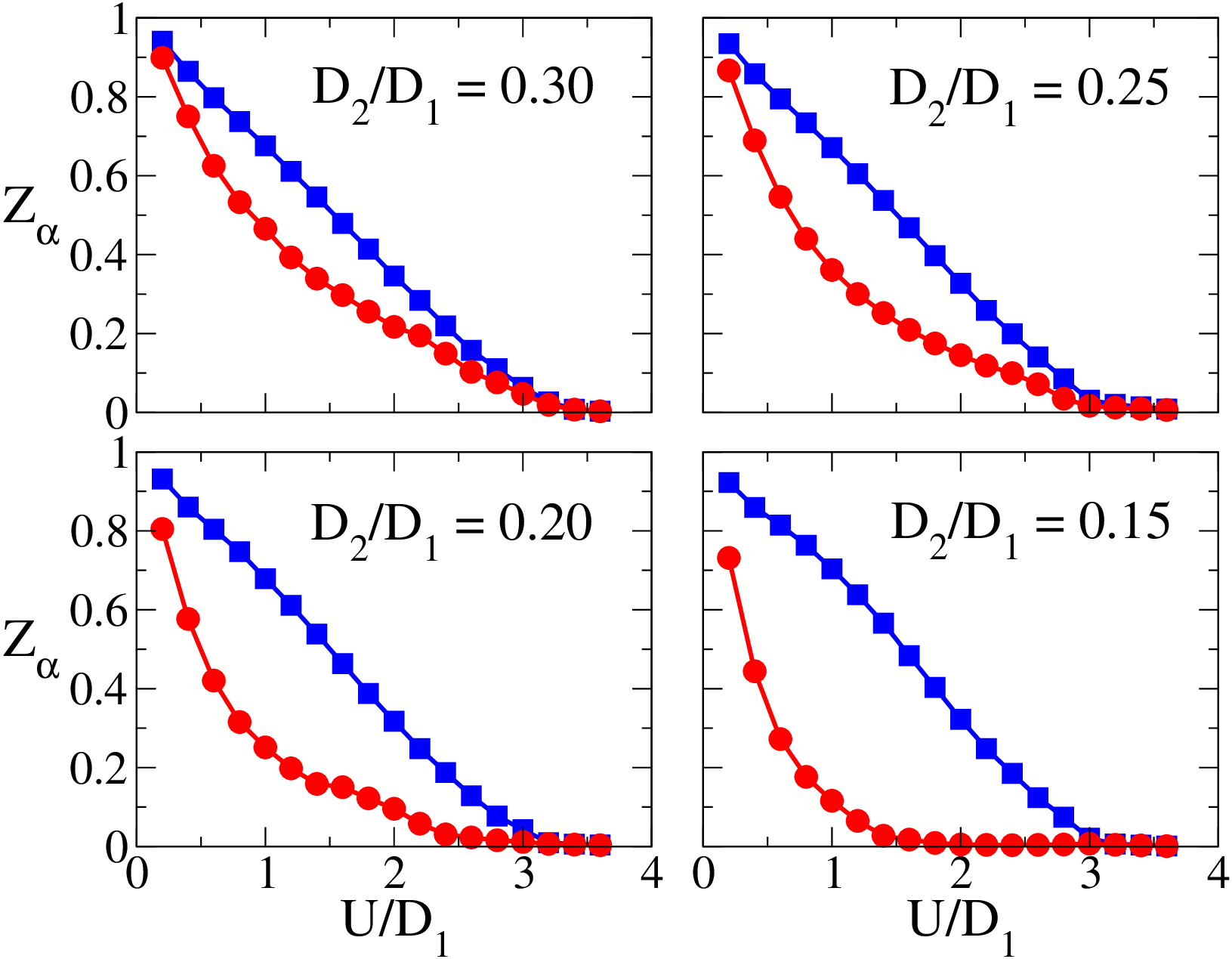}
\caption{
  Quasiparticle residues $Z_a$ for the wide band (squares) and for the
  narrow band (circles) obtained by the DMFT calculation for different
  ratios of the bandwidth $D_2 / D_1$ and $J = 0$.}
\label{fg:ZDMFTJ000}
\end{figure}
%%%%%%%%%%%%%%%%%%%%%%%%%%%%%%%%%%%%%%%%%%%%%%%%%%%%%%%%%%%%%%%%%%%%%%%%%%%

First, we treat the case in which there is no Hund's coupling (i.e.,
$J=0$). The metallic or insulating nature of each band is
characterized by its quasiparticle residue $Z_a^{-1} = 1 -
\left(\partial \mathcal{I}m \Sigma_a(i \omega) / \partial i
  \omega\right)\lvert_{\omega \rightarrow 0}$. In
Fig.~\ref{fg:ZDMFTJ000}, we show $Z_a$ as a function of the Coulomb
repulsion $U$ for different bandwidth ratios $D_2 / D_1$.  When the
ratio of the bandwidth $D_2 / D_1 \ge 0.20$, the quasiparticle weights
decrease as the Coulomb interaction gets bigger.  Even though there is
a stronger initial reduction for the narrow band, the weights
eventually vanish for the same critical value of $U_c/D_1 \simeq 3.6$.
The situation changes when we further decrease the bandwidth ratio
(i.e., $D_2 / D_1 = 0.15$). In this case, we find that the weights of
the bands vanish for different values of $U$, in agreement with the
results of the Gutzwiller wave function.  Moreover, the critical ratio
$D_2 / D_1 = 0.20$ that we found earlier seems consistent with the
DMFT calculation.

%%%%%%%%%%%%%%%%%%%%%%%%%%%%%%%%%%%%%%%%%%%%%%%%%%%%%%%%%%%%%%%%%%%%%%%%%%%
\begin{figure}
\includegraphics[width=0.48\textwidth]{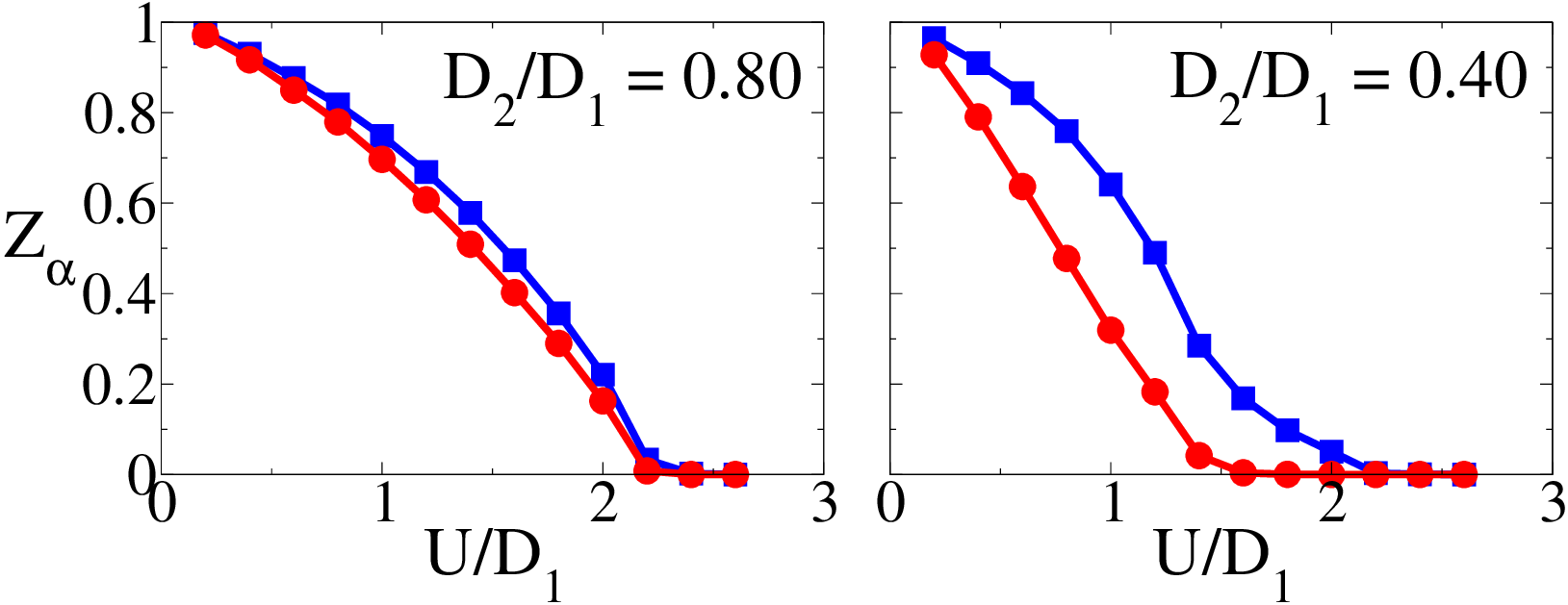}
\caption{
  Quasiparticle residues $Z_a$ for the wide band (squares) and the
  narrow band (circles) obtained by the DMFT calculation for different
  ratios of the bandwidth $D_2 / D_1$ and $J / U = 0.10$.}
\label{fg:ZDMFTJ010}
\end{figure}
%%%%%%%%%%%%%%%%%%%%%%%%%%%%%%%%%%%%%%%%%%%%%%%%%%%%%%%%%%%%%%%%%%%%%%%%%%%

Let us turn now to the model in the presence of a finite Hund's
coupling, $J / U = 0.10$, and perform the same calculations for the
following ratios of the bandwidths: $D_2 / D_1 = 0.80$ and $0.40$. The
results are shown in Fig.~\ref{fg:ZDMFTJ010}. We still find evidences
for an OSMT, this time, however, below a larger ratio of the
bandwidths.  Further calculations show that the critical ratio of the
bandwidth for $J / U = 0.10$ is $D_2 / D_1 \sim 0.60$.

If we were to confine our analysis to the behavior of the
quasiparticle residues $Z_a$'s, we should conclude that, both in the
absence and in the presence of a Hund's coupling, the OSMT scenario
does occur for sufficiently small ratio of the bandwidth, in
qualitative and also quantitative agreement with the variational
results of the Gutzwiller wave function. The only difference may lie in
the order of the transition. Even if it is difficult to settle
precisely the order of the transition within DMFT, our results seem to
point towards a second-order phase transition at the MIT $U_1$ with
finite $J$, contrary to the first-order transition predicted by the
Gutzwiller wave function.

%%%%%%%%%%%%%%%%%%%%%%%%%%%%%%%%%%%%%%%%%%%%%%%%%%%%%%%%%%%%%%%%%%%%%%%%%%%
\begin{figure}
\includegraphics[width=0.48\textwidth]{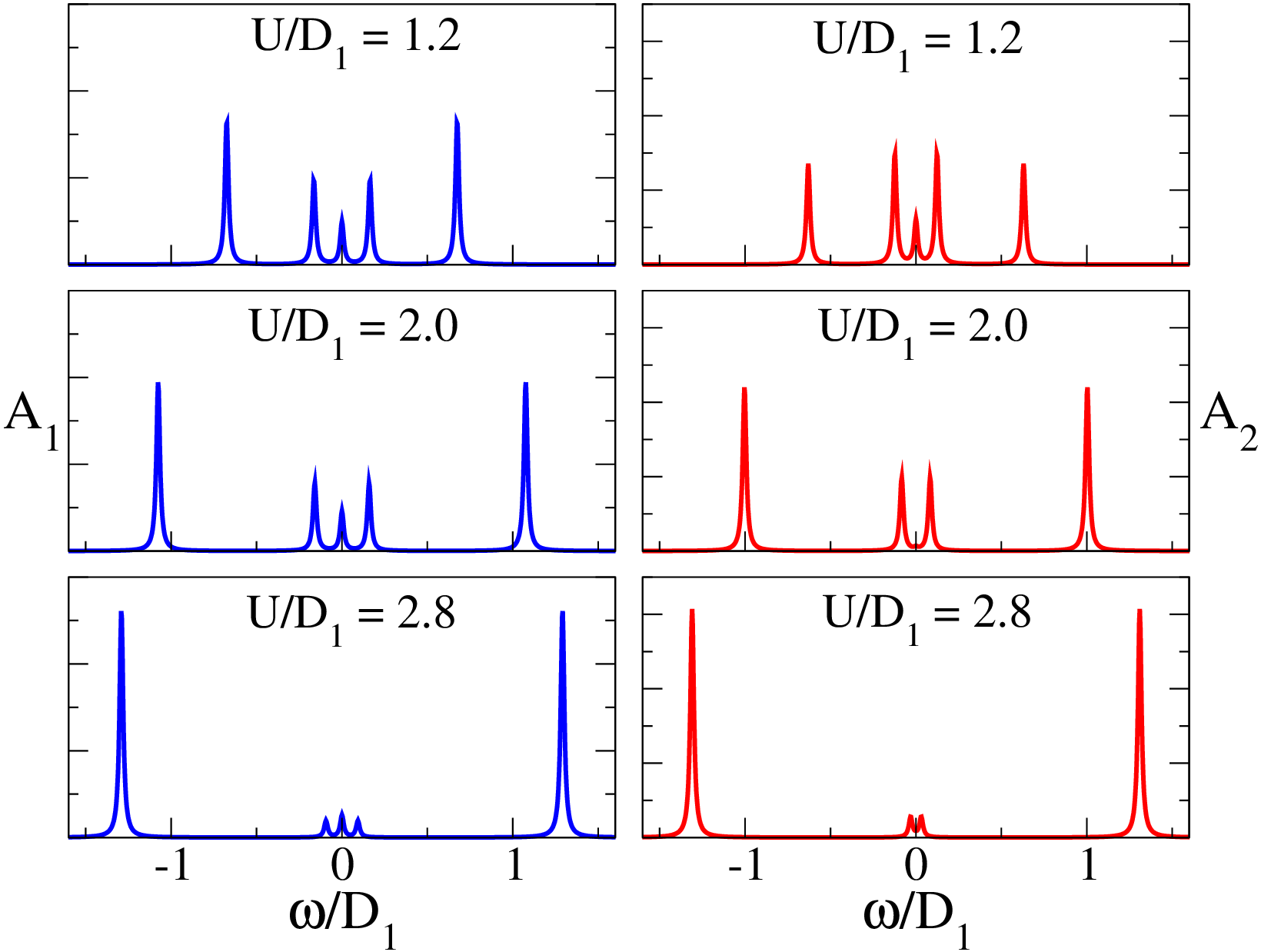}
\caption{
  Density of states for the wide band (left panels) and the narrow
  band (right panels) as obtained by DMFT for different values of the
  Coulomb interaction. The ratio of the bandwidth is $D_2/D_1 = 0.15$
  and $J = 0$.}
\label{fg:density-DMFT-U}
\end{figure}
%%%%%%%%%%%%%%%%%%%%%%%%%%%%%%%%%%%%%%%%%%%%%%%%%%%%%%%%%%%%%%%%%%%%%%%%%%%

%%%%%%%%%%%%%%%%%%%%%%%%%%%%%%%%%%%%%%%%%%%%%%%%%%%%%%%%%%%%%%%%%%%%%%%%%%%
\begin{figure}[!ht]
\includegraphics[width=0.48\textwidth]{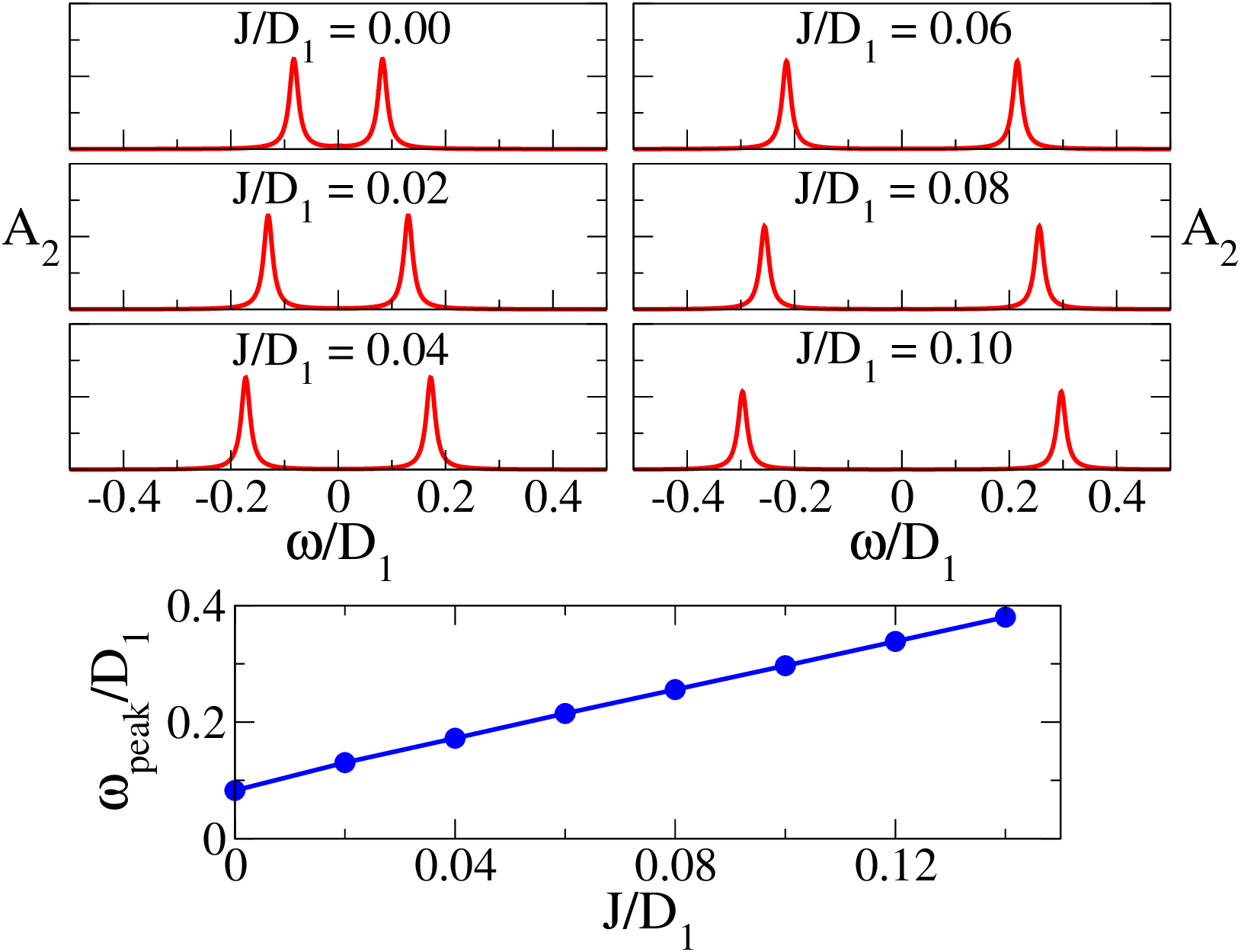}
\caption{
  Upper panels: low-energy part of the density of states of the narrow
  band as obtained by DMFT for different values of $J/D_1$. The ratio
  of the bandwidths is $D_2 / D_1 = 0.15$, and the Coulomb repulsion
  is set to $U/D_1 = 2$.  Lower panel: position of the low-energy
  peaks as a function of $J/D_1$.  }
\label{fg:density-DMFT-J}
\end{figure}
%%%%%%%%%%%%%%%%%%%%%%%%%%%%%%%%%%%%%%%%%%%%%%%%%%%%%%%%%%%%%%%%%%%%%%%%%%%

A deeper insight into the above scenario can be gained by analyzing 
the spectral properties of the more correlated band, and not just its 
quasiparticle residue. Indeed, such an inspection leads to a less clear-cut
picture, revealing features which are not captured by the Gutzwiller
wave function. In Fig.~\ref{fg:density-DMFT-U}, we show the density of states (DOS) 
of both orbitals for various Hubbard $U$'s and $D_2 / D_1 = 0.15$.  We
notice that, although the DOS of the narrow band right at the chemical
potential becomes zero within our numerical accuracy above $U_2$,
there is still low-energy spectral weight inside the Mott-Hubbard gap.
Due to our discretization procedure, this weight is concentrated in
two peaks located symmetrically with respect to the chemical
potential.  These peaks are also present at $J=0$, and move linearly away
from the chemical potential when $J\not =0$, roughly as $2J$, see
Fig.~\ref{fg:density-DMFT-J}. Their total spectral weight scales
approximately like the quasiparticle residue of the wider band, $Z_1$,
both vanishing at the second MIT, $U_1$. 
In addition, if $J=0$,  the distance between the peaks also scales
like $Z_1$, see Fig.~\ref{fg:density-DMFT-U}. If we, reasonably,
assume that these two peaks mimic two resonances, one below and the
other above the chemical potential, it becomes much less obvious what
might be the actual value of the DOS right at the chemical potential
if we were not constrained to a small number of levels.  Moreover, even
if the DOS were strictly zero at the chemical potential, still we
should determine whether band 2 behaves like a small-gap semiconductor
or a semimetal for $U_2\leq U \leq U_1$. In other words, the energy
discretization inherent in the exact diagonalization technique might
play a more critical role in our case than in the simplest single-band
Hubbard model.

Therefore, although the numerical evidences we have presented so far
point in favor of the existence of an OSMT with zero or finite $J$
below a critical bandwidth ratio, there are several aspects which 
still need to be clarified. We will consider a deeper investigation of
such aspects in the following sections.

\section{Single impurity spectral properties}\label{sc:AIM}

The first issue we want to address concerns the origin of the two
peaks in the orbital 2 spectral function inside the Mott-Hubbard gap.
Although the self-consistency condition~(\ref{eq:selfcons}) of the
effective Anderson impurity model~(\ref{eq:Hand}) plays a very crucial
role, for instance it determines a critical value of $U$ above which the Kondo
effect does not take place anymore, yet useful information can be obtained by
studying~(\ref{eq:Hand}) without imposing~(\ref{eq:selfcons}), which
is what we are going to do in this section by means of the Wilson
Numerical Renormalization Group (NRG).~\cite{NRG}

\begin{figure}
\includegraphics[width=0.48\textwidth]{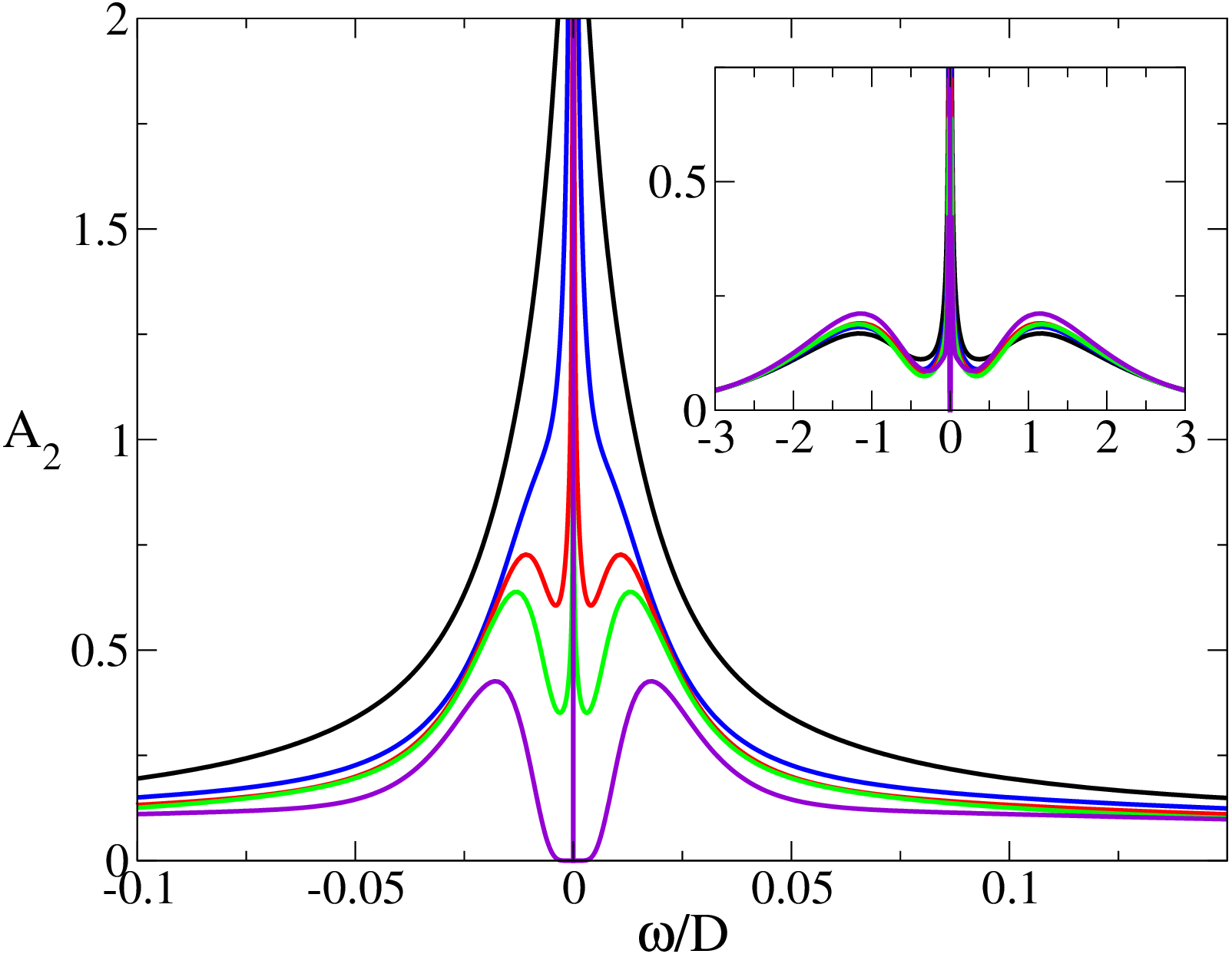}
\caption{
  Spectral function $A_2 (\omega)$ of the narrow band for $J = 0$, $U
  / D = 2$, $\Gamma_1 = D / 2 \pi$, and $\Gamma_2 / \Gamma_1 =
  \frac{1}{2}, \frac{1}{4}, \frac{1}{6}, \frac{1}{8}, 0$ (from top to
  bottom). Inset: the same spectral function on a wider scale.}
\label{fg:density-NRG-U}
\end{figure}

\begin{figure}
\includegraphics[width=0.48\textwidth]{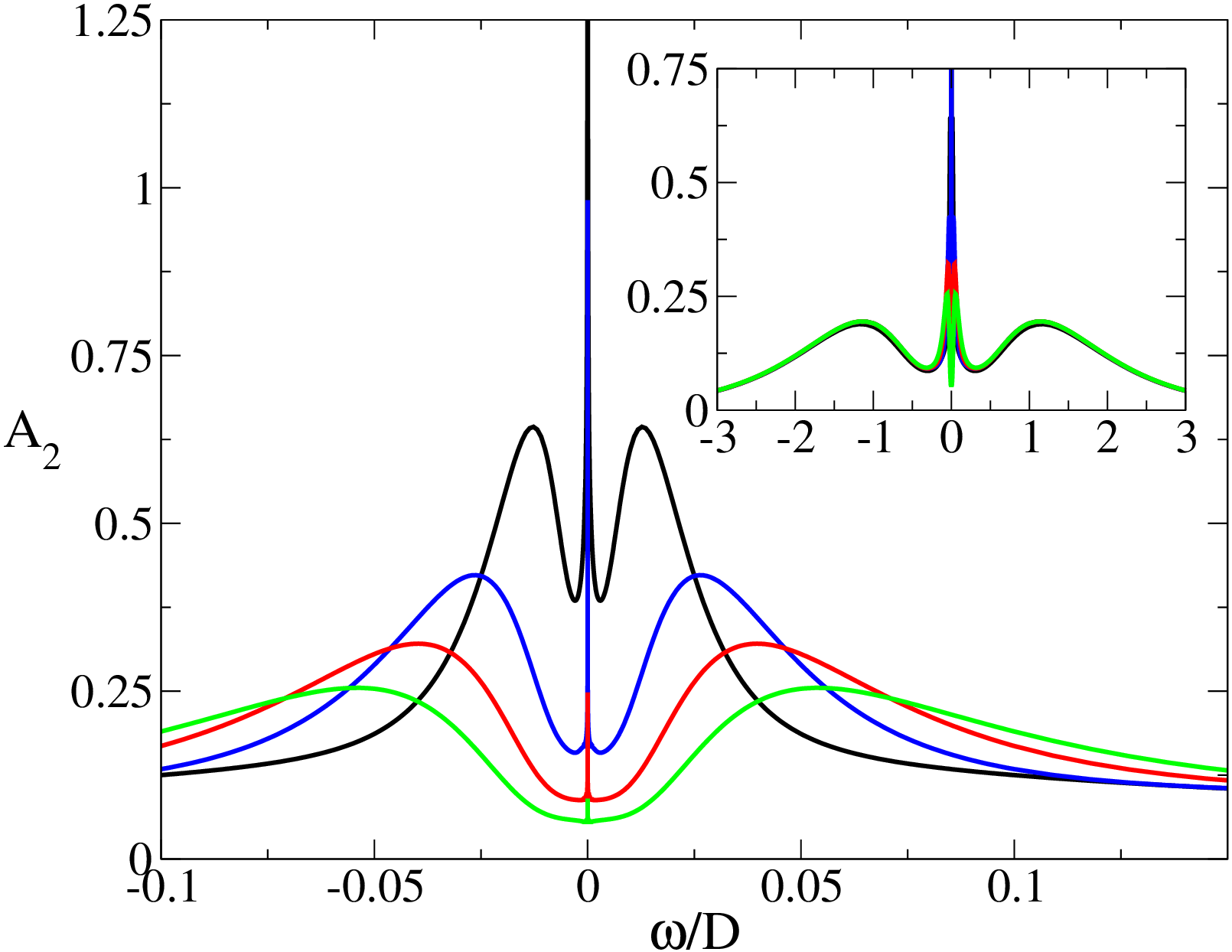}
\caption{
  Spectral function $A_2 (\omega)$ of the narrow band for $J/D = 0,
  0.004, 0.008, 0.012$ (from top to bottom), $U / D = 2$, 
  $\Gamma_1 = D / 2 \pi$, and $\Gamma_2 / \Gamma_1 = 1/8$. 
  Inset: the same spectral function on a wider scale.}
\label{fg:density-NRG-J}
\end{figure}

The Anderson impurity model~(\ref{eq:Hand}) is controlled by several
energy scales, the Hubbard $U$, the Hund's coupling $J$ and the
so-called hybridization widths
\[
\Gamma_a = \sum_k \, V_{k a}^2\, \delta\left(\epsilon_{ka}\right).
\]
For simplicity, we will assume that the two conduction baths are
degenerate with half-bandwidth $D$, which will be our unit of energy.
In Fig.~\ref{fg:density-NRG-U}, we show the impurity spectral function
of the orbital 2, $A_2(\omega)$, as obtained by NRG for $J=0$, $U=2D$,
$\Gamma_1=D/(2\pi)$, and for several values of $\Gamma_2/\Gamma_1<1$.
Since we do not impose any self-consistency, the DOS for any
$\Gamma_2\not = 0$ shows a Kondo resonance at the chemical potential,
which narrows as $\Gamma_2$ decreases. In addition, there are two more
peaks which move slightly away from the chemical potential as
$\Gamma_2$ is reduced. These peaks actually resemble those we find in
the DMFT calculation. Indeed, they move linearly as we switch on $J$,
see Fig.~\ref{fg:density-NRG-J}, just like we observe within DMFT.

%%%%%%%%%%%%%%%%%%%%%%%%%%%%%%%%%%%%%%%%%%%%%%%%%%%%%%%%%%%%%%%%%%%%%%%%%%%
%%%%%%%%%%%%%%%%%%%%%%%%%%%%%%%%%%%%%%%%%%%%%%%%%%%%%%%%%%%%%%%%%%%%%%%%%%%

%%%%%%%%%%%%%%%%%%%%%%%%%%%%%%%%%%%%%%%%%%%%%%%%%%%%%%%%%%%%%%%%%%%%%%%%%%%
%%%%%%%%%%%%%%%%%%%%%%%%%%%%%%%%%%%%%%%%%%%%%%%%%%%%%%%%%%%%%%%%%%%%%%%%%%%

The origin of these peaks is easy to identify when $\Gamma_2=0$.  When
the orbital 2 is not hybridized with its bath, its occupation number
$n_2$ is a conserved quantity. The ground state is expected to belong
to the subspace with $n_2=1$, because, in this case, $\langle n_1
\rangle = 1$ and the Kondo-screening energy gain is maximum. This
state is twofold degenerate reflecting the free spin-1/2 of the
electron localized in orbital 2.  The energy gap to the lowest energy
states for $n_2=0,2$, $\langle n_1\rangle = 2,0$, respectively, is
therefore of the order of the Kondo temperature of orbital 1, 
$T_{K\, 1}$. The impurity orbital-2 DOS is analogous to the core-hole
spectral function in X-ray absorption, so it should start above a
finite threshold proportional to $T_{K\, 1}$. In other words, the DOS
has a small but finite gap of order $T_{K\, 1}$, similar to what we
observe within DMFT.

However, as soon as $\Gamma_2$ is non zero, this gap is filled and, in
addition, a Kondo-resonance appears. Even if we move the peaks away
from the chemical potential by increasing $J$, see
Fig.~\ref{fg:density-NRG-J}, the region between them and the narrow 
(practically invisible in the figure) Kondo-resonance is still covered by 
spectral weight. In the light of this dynamical behavior, it is not at all 
obvious what the self-consistency requirement~(\ref{eq:selfcons})  
may lead to when band-1 is still conducting. In other words, either 
a true narrow gap, as if $\Gamma_2=0$, or a pseudo-gap with a power-law vanishing DOS, 
or two-peaks plus the narrow resonance are equally compatible with 
the self-consistency condition.    
However, the event in which most of the spectral weight is concentrated in the 
two symmetric peaks, leaving only 
negligible weight within the narrow resonance, is extremely hard to identify 
with a limited number of levels. As an attempt to discriminate among the aforementioned 
possible scenarios, in the following section, we implement a projective self-consistency 
technique which allows a more detailed low-energy description within DMFT.

\section{Projective Self-Consistent Technique}\label{sc:psct}

A remarkable feature uncovered by DMFT nearby a MIT is the clear
separation of energy scales between well preformed high-energy
Hubbard bands and lingering low-energy itinerant quasiparticles. It
has been shown,~\cite{moeller1995} that this partition of energy scales
allows to reformulate the problem into a new one, in which the
high-energy part is projected out. Essentially, the original Anderson
impurity model, which involves both high-energy side-bands and
low-energy quasiparticles, is reduced to a Kondo-like model which can
be attacked more easily by a numerical procedure. In this section,
we apply a projective technique to our model, which is similar to
Ref.~\onlinecite{moeller1995}, with the only difference that the
resulting effective problem is still an Anderson impurity model
with rescaled parameters.

As we showed, the occurrence of an OSMT does not seem to require a
finite exchange but rather a sufficiently small bandwidth ratio.
Therefore, we prefer to present the projective technique in the simpler
case where $J=0$. Following Ref.~\onlinecite{moeller1995}, we start by
rewriting the Anderson impurity model~(\ref{eq:Hand}) explicitly
separating low- ($L$) and high- ($H$) energy scales
\be\label{eq:Hfull}
\mathcal{H}_\mathrm{AM} = \mathcal{H}^H + \mathcal{H}^L + \mathcal{H}^m,
\ee
where
\begin{multline}\label{eq:Hhigh}
  \mathcal{H}^H =
    \frac{U}{2} (n_f - 2)^2 \\
    + \sideset{}{^H}\sum_{k, a, \sigma}
    \epsilon_{k a}^H
    c^\dagger_{k a \sigma} c^{\phantom{\dagger}}_{k a \sigma}
    + \sideset{}{^H}\sum_{k, a, \sigma}
    V_{k a}^H
    (f^\dagger_{a \sigma} c^{\phantom{\dagger}}_{k a \sigma}
    + h.c.), 
\end{multline}
describes the impurity coupled to the high-energy levels, 
\be\label{eq:Hlow}
    \mathcal{H}^L= \sideset{}{^L}\sum_{k, a, \sigma}
    \epsilon_{k a}^L
    c^\dagger_{k a \sigma} c^{\phantom{\dagger}}_{k a \sigma},
\ee
is the low-energy bath Hamiltonian, and finally  
\be\label{eq:Hmix}
\mathcal{H}^m = 
     \sideset{}{^L}\sum_{k, a, \sigma}
    V_{k a}^L
    (f^\dagger_{a \sigma} c^{\phantom{\dagger}}_{k a \sigma}
    + h.c.),
\ee
mixes low- and high-energy sectors. The impurity Green's function,
also written as sum of a low- and a high-energy part, $G_a(i \omega) =
G_a^L(i \omega) + G_a^H(i \omega)$, should satisfy the
self-consistency requirement~(\ref{eq:selfcons}). If we assume that
the low-energy spectral weight is $W_a\ll 1$, the self-consistency
condition for the integrated low- and high-energy spectral functions,
$\rho^L_a(\epsilon)$ and $\rho^H_a(\epsilon)$, respectively, implies
the following sum-rules
\ba
\sideset{}{^L}\sum_{k}\, \left(V_{k a}^L\right)^2 &=& 
t_a^2\,W_a,\label{Low-int}\\
\sideset{}{^H}\sum_{k}\, \left(V_{k a}^H\right)^2 &=& 
t_a^2\,\left(1-W_a\right),\label{High-int}
\ea
showing that the impurity is strongly hybridized with the high-energy
levels and very weakly
with the low-energy ones. Let us for the moment neglect the coupling
to the latter. The ground state of~(\ref{eq:Hhigh}) is
the adiabatic evolution of the states in which all negative-energy
bath levels are doubly occupied and two electrons sit on the impurity,
giving rise to a six-fold degenerate ground state.
Other states with the same number of electrons lie
above the ground state at least by an energy $U$. The lowest-energy
states with one more (less) electron are more degenerate, since they
emerge adiabatically from the states obtained by adding (removing) an
electron either in the impurity levels or in the positive (negative)-energy
baths. This large degeneracy is, however, split linearly by $V_{k a}^H$,
which implies the broadening of the Hubbard bands around their
centers of gravity $\pm U/2$. The main effect of the mixing 
term~(\ref{eq:Hmix}) is to provide a Kondo exchange coupling between the
six-fold degenerate ground state of $\mathcal{H}^H$ and the low-energy
baths, which can be obtained by degenerate second-order perturbation
theory in $\mathcal{H}^m$ or, more formally, by a Schrieffer-Wolff
canonical transformation.~\cite{Schrieffer,moeller1995} Once the
effective Kondo model is obtained, we could for instance follow
Ref.~\onlinecite{moeller1995}, namely solve that model and impose the
self-consistency condition to the impurity Green's function,
calculated through the Schrieffer-Wolff canonically transformed
$f_{a\sigma}$. To be consistent, one should in principle expand the
transformed $f_{a\sigma}$ up to second order in $V^L/U$ and impose the
self-consistency requirement in the whole energy range, including low
and high energies. In practice, even if the self-consistency is imposed
only to the low-energy spectrum, one still gets a faithful description
of the critical behavior near the MIT.~\cite{moeller1995}
An equivalent procedure, that we have instead decided to follow, is to
identify a new two-orbital Anderson impurity model, coupled only to the
low-energy levels, which maps to the same Kondo model and next impose
the self-consistency only to the low-energy part of the impurity
Green's function:
\be\label{L-self consistency}
t_a^2\, G^L_a(i\omega_n) = 
\sum_k\, \frac{\left(V_{k a}^L\right)^2}{i\omega_n-\epsilon^L_{k a}}.
\ee
Regarding the high-energy part of the self-consistency, since we
always model the high-energy levels with just four levels at energies
$\epsilon^H_{a\pm} = \pm U/2$, we need to impose an additional
requirement besides~(\ref{L-self consistency}), which, through~(\ref{High-int}),
is simply
\be
V^H_{a\pm} = t_a\, \sqrt{\frac{1-W_a}{2}},
\label{additional}
\ee
where $W_a$ is the low-energy spectral weight obtained self-consistently
from~(\ref{L-self consistency}). The advantage of the projective method is
that we can now model the low-energy spectrum with more levels, 
the cost being the additional self-consistency condition~(\ref{additional}).

\begin{figure}
\includegraphics[width=0.48\textwidth]{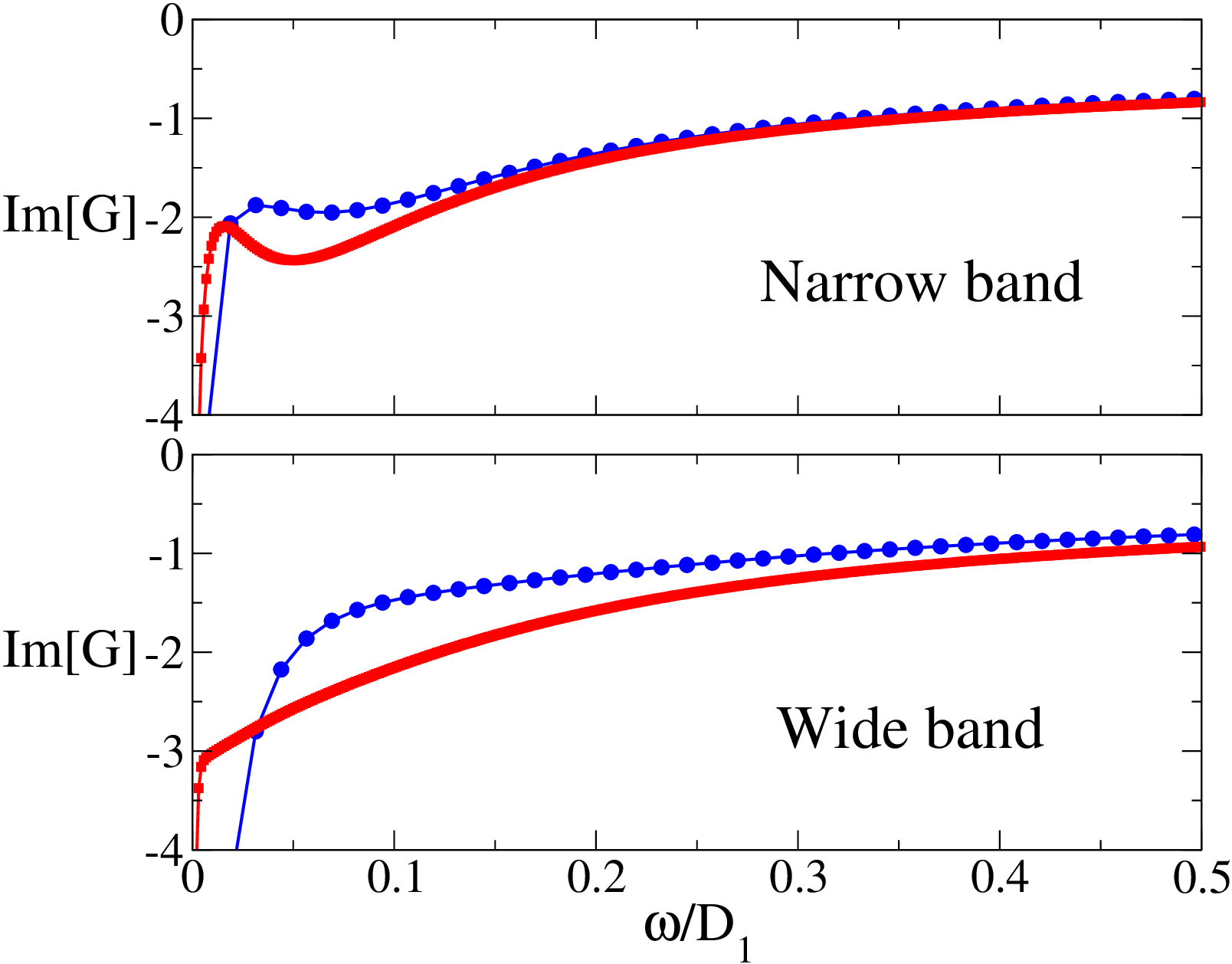}
\caption{
  Imaginary part of the Green's function for $D_2 / D_1 = 0.18$, $U /
  D_1 = 2.00$, and $J = 0$ as obtained by DMFT and with the PSCT.}
\label{fg:greencomparison}
\end{figure}

When we apply this projective technique to our two-orbital model with
$J=0$ and $t_1\geq t_2$, we end up with an effective Anderson impurity
model
\begin{multline}\label{eq:Heff}
  \mathcal{H}_\mathrm{eff} =
    \frac{U_1}{2} (n_1 - 1)^2 + \frac{U_2}{2} (n_2 - 1)^2
    + \frac{U_{12}}{2} (n_1 - 1)(n_2 - 1) \\
    + \sideset{}{^L}\sum_{k, a, \sigma}
    \epsilon_{k a}^L
    c^\dagger_{k a \sigma} c^{\phantom{\dagger}}_{k a \sigma}
    + \sideset{}{^L}\sum_{k, a, \sigma}\, V_{k a}^L
    (f^\dagger_{a \sigma} c^{\phantom{\dagger}}_{k a \sigma}
    + h.c.),
\end{multline}
where $U_1$, and $U_2$ are found from the solution of the
high-energy problem~(\ref{eq:Hhigh}). Moreover,
$U_{12}=\left(U_1+U_2\right)/2$, which assures the six-fold
degeneracy of the isolated doubly-occupied impurity and we have that
$U_1\leq U_2$ with $U_2-U_1\sim t_1-t_2$. In other words, the high-energy levels
provide a partial screening of the Hubbard repulsion, more efficient
within orbital 1, which is more hybridized with the bath. Therefore, the
difference of bandwidths acquires in our projective method a quite
transparent role: while the {\it bare} Coulomb repulsion does not care
about the orbitals in which electrons sit, this indifference is lost
once the high-energy screening is taken into account. In
Fig.~\ref{fg:greencomparison}, we compare the imaginary part of the
Green's functions in Matsubara frequencies $\omega_n$ as function of
$\omega_n$ as obtained by full DMFT or using the above projective
self-consistent technique (PSCT), at $J=0$. Note that within the PSCT, 
we model the low-energy conduction bath through five discrete levels.
The agreement is satisfying, and the additional levels clearly allow
for a more accurate description of the low-energy Green's function.

%%%%%%%%%%%%%%%%%%%%%%%%%%%%%%%%%%%%%%%%%%%%%%%%%%%%%%%%%%%%%%%%%%%%%%%%%%%
%%%%%%%%%%%%%%%%%%%%%%%%%%%%%%%%%%%%%%%%%%%%%%%%%%%%%%%%%%%%%%%%%%%%%%%%%%%

%%%%%%%%%%%%%%%%%%%%%%%%%%%%%%%%%%%%%%%%%%%%%%%%%%%%%%%%%%%%%%%%%%%%%%%%%%%
\begin{figure}
\includegraphics[width=0.48\textwidth]{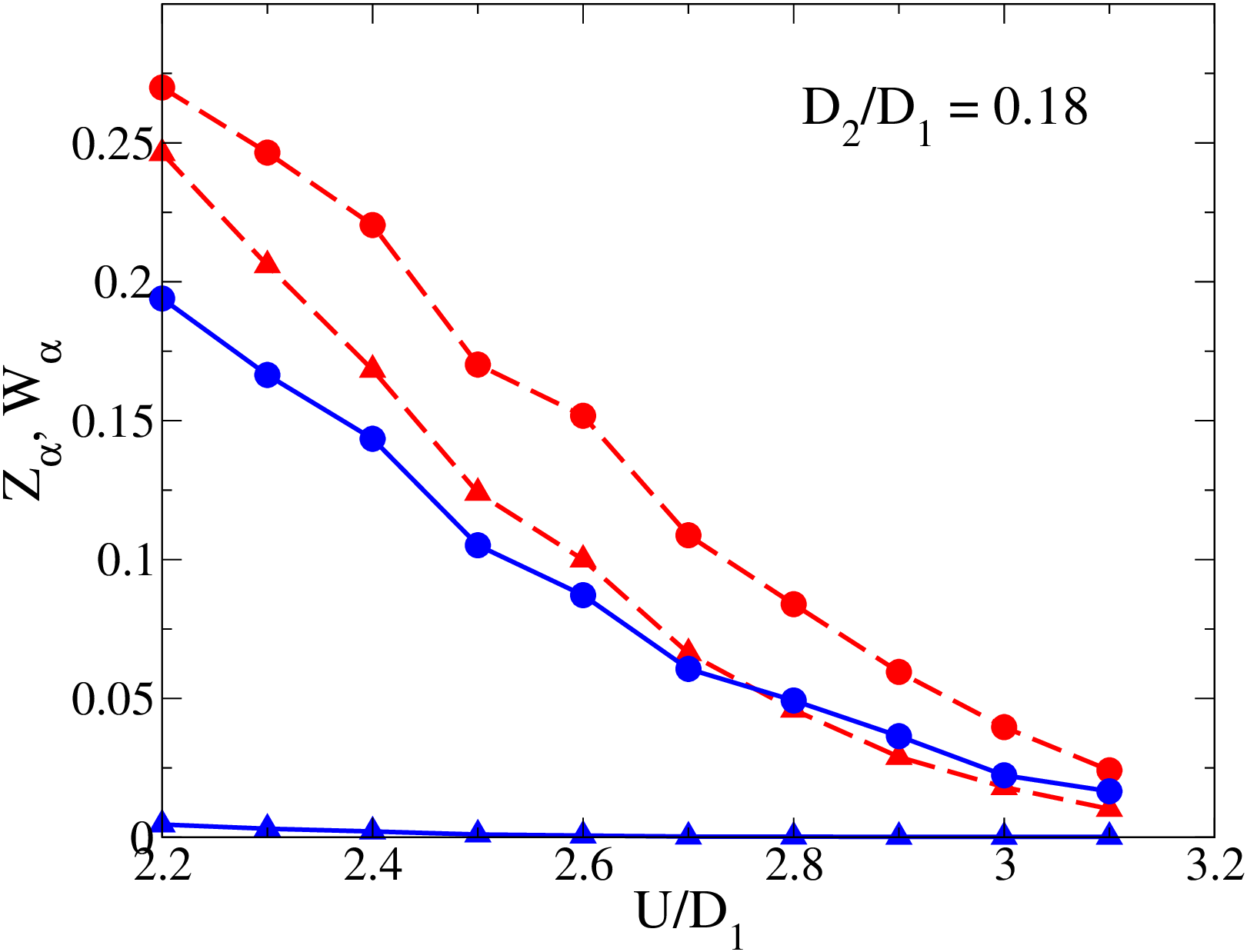}
\caption{
  Quasiparticle residues $Z_a$ (triangles) and low-energy spectral
  weights $W_a$ (circles) as obtained by the PSCT with $J = 0$. Dashed
  lines refer to the wide band, and continuous lines to the narrow
  one.}
\label{fg:ZW-PSCT}
\end{figure}
%%%%%%%%%%%%%%%%%%%%%%%%%%%%%%%%%%%%%%%%%%%%%%%%%%%%%%%%%%%%%%%%%%%%%%%%%%%

%%%%%%%%%%%%%%%%%%%%%%%%%%%%%%%%%%%%%%%%%%%%%%%%%%%%%%%%%%%%%%%%%%%%%%%%%%%
\begin{figure}
\includegraphics[width=0.48\textwidth]{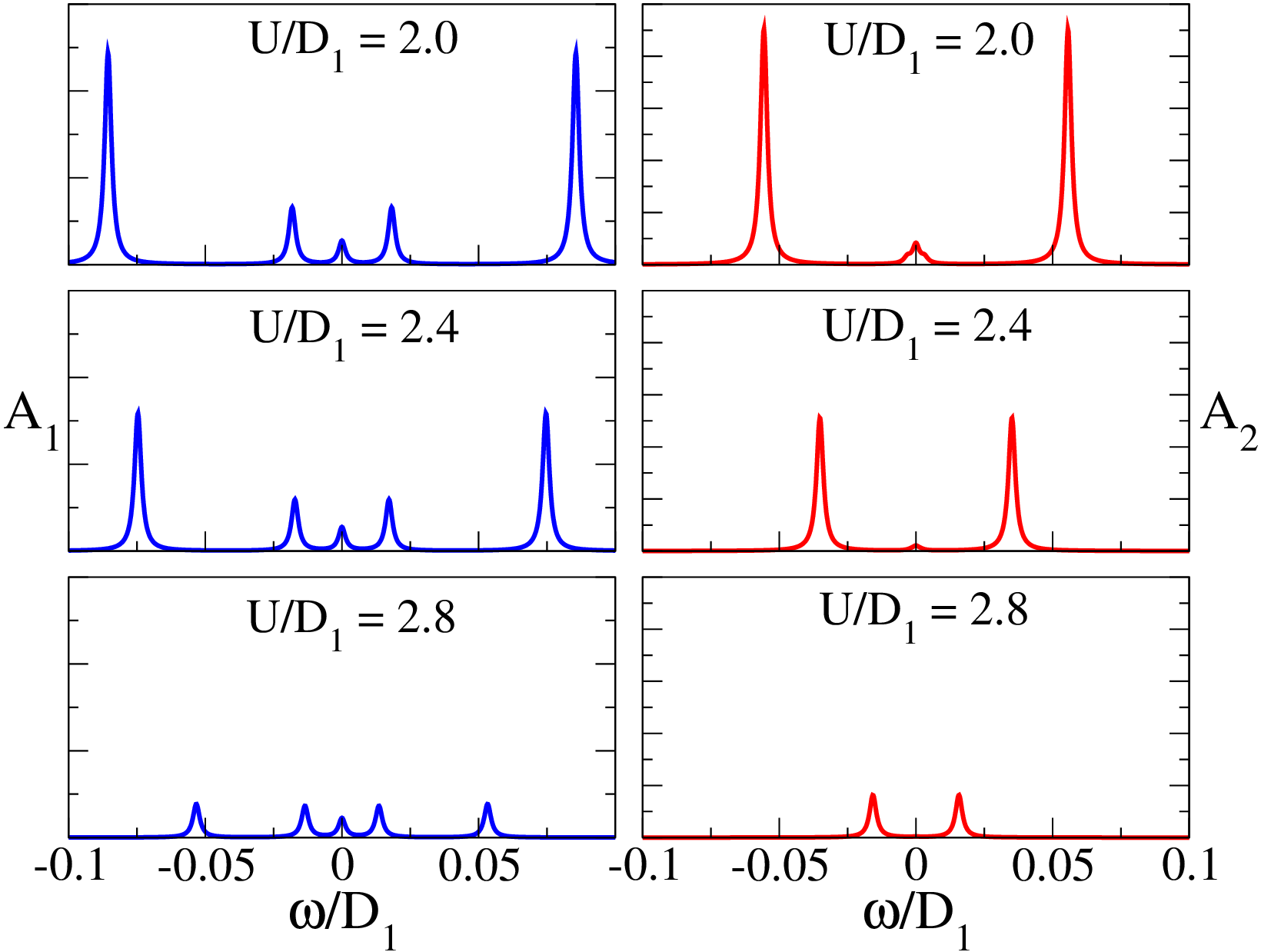}
\caption{
  Low-energy part of the density of states of the wide band (left
  panels) and the narrow band (right panels) obtained within the PSCT
  for different $U / D_1$. The ratio of the bandwidth is $D_2 / D_1 =
  0.18$ and $J = 0$.}
\label{fg:density-PSCT}
\end{figure}
%%%%%%%%%%%%%%%%%%%%%%%%%%%%%%%%%%%%%%%%%%%%%%%%%%%%%%%%%%%%%%%%%%%%%%%%%%%

In Fig.~\ref{fg:ZW-PSCT}, we plot the PSCT values of the quasiparticle
residues $Z_1$ and $Z_2$ as function of $U$ at $J=0$ for
$D_2/D_1=0.18$, as well as of the full spectral weights, $W_1$
and $W_2$, inside the Mott Hubbard gap. In agreement with standard DMFT,
we find a region where $Z_2$ is zero within our numerical accuracy,
while $Z_1$ is still finite. Yet, the total spectral weights are both
non zero. In Fig.~\ref{fg:density-PSCT}, we draw the low-energy DOS
for the two bands and various $U$'s.

\section{Discussion and conclusions}\label{sc:conclusion}

In this work, we have studied by several techniques the properties of the Mott 
transition in an infinite-dimensional Hubbard model with two bands having the same 
center of gravity but different bandwidths, both in the presence and in the absence of the 
Hund's exchange splitting $J$. 
We have shown that a variational calculation based on a Gutzwiller wave function predicts 
that the two bands may undergo different metal-insulator transitions both for 
$J=0$ and $J\not = 0$: by increasing $U$, the narrower band 
ceases to conduct before the wider one.
The necessary condition for this orbital-selective Mott transition is that the 
bandwidth ratio is lower than a critical value which increases with $J$,  
being 0.2 when $J=0$, see Fig.~\ref{fg:phasediag}. This result would contradict both 
Liebsch,~\cite{liebsch2003,liebsch2004} who claims that there is only a single 
Mott transition whatever is $J$ and the bandwidth ratio, as well as  
Koga and coworkers,~\cite{koga2004} who instead suggest that 
the Mott transition is unique if $J=0$ but splits into two distinct ones as soon as 
$J\not = 0$ for any value of the bandwidth ratio.
The behavior of the quasiparticle residues as obtained by DMFT using exact diagonalization 
confirms, even quantitatively, the variational results, showing that the 
residue of the narrower band may vanish before 
the one of the wider band if the bandwidth ratio is sufficiently small, both 
for $J=0$ and $J\not =0$. Actually, an OSMT which occurs both in the absence and in 
the presence of an exchange splitting is somehow more conceivable, because  
the role of $J$ is model dependent.
We notice that, in more general situations where the number
of orbitals is greater than two and different from the number of
electrons, as for instance in the case of $t_{2g}$ orbitals occupied
by two or four electrons on average, the Coulomb exchange would
instead compete against the angular momentum quenching which occurs in
the OSMT scenario. Therefore, we suspect that the role of the Coulomb
exchange might actually depend on the specific model.

However, a closer inspection to the low-energy spectral properties of the 
narrower band in the region where it is apparently insulating while the wider band 
still conducts poses doubts to the above simple scenario. The reason is that, in spite 
of a quasiparticle residue which is zero within our numerical accuracy, the narrower band 
has spectral weight inside the Mott-Hubbard gap, which scales 
like the quasiparticle residue of the wider band. In other words,
the charge fluctuations which still occur in the wider band are transferred 
into the narrower one, as somehow predictable. This low-energy spectral weight is 
concentrated in two peaks symmetrically located around the chemical potential. 
Roughly speaking, the distance of each peak 
from the chemical potential is $2J$ plus a quantity of the order of the quasiparticle 
resonance width of the wider band. 
Due to our limited numerical resolution, we can 
not establish rigorously whether these two peaks 
(a) signal a narrow-gap semiconducting behavior,
(b) signal a semimetallic behavior, with a power-law vanishing density of states,
(c) or coexist with an extremely narrow resonance at the chemical potential,  
with a spectral weight well below our numerical accuracy, just like the 
single impurity does. Although 
the elements at our disposal do not definitely allow to discriminate among these  
three scenarios, yet one can recognize that some of them are more plausible than the others. 

The first possibility (a) of a narrow-gap semiconductor seems very unlikely. 
Indeed, in this case, the gap between the two low-energy peaks would open large and then
diminish as the quasiparticle resonance width of the wider band, by further 
increasing the repulsion $U$. Therefore, the insulating character of the narrower band 
would weaken by increasing $U$, which seems a bit odd. 

Let us consider instead the scenario (b) of a semimetal. If taken literally, 
it would imply a vanishingly small local magnetic susceptibility, while we 
actually find a very large one, much larger than the local susceptibility of the wider band. 
However, a semimetallic behavior would imply, in our particle-hole symmetric case, 
a breakdown of Fermi liquid theory.~\cite{nota} 
Therefore, a power-law vanishing single-particle DOS might not necessarily conflict with 
almost free-spin excitations in a scenario in which Fermi liquid theory breaks down 
and for instance spin-charge separation emerges. Although it might represent a 
quite interesting circumstance, yet we could not find any physical 
arguments justifying such a non-Fermi liquid behavior.
Therefore, we are tempted to discard it 
in favor of the more conservative scenario (c) in which the two peaks 
coexist with a narrow resonance which remains tied 
at the chemical potential, its spectral weight being smaller than 
our numerical accuracy. This resonance should disappear right at the same $U$ 
where the wider band ceases to conduct.
 
In any case, the rich structure of the low-energy spectrum, revealed by our study,
highlights the subtlety of the present problem and could explain
the discrepancies between previous studies. 
Further investigation is needed to confirm our speculation that 
the narrow resonance peak displayed by the single impurity 
survives the DMFT self-consistency.

\acknowledgments

During the completion of this paper, we have learned about the work by L.  
de' Medici, A. Georges, and S. Biermann, that leads to similar
conclusions. We acknowledge helpful discussions with C. Castellani, L.  
de' Medici, A. Georges, A.  Koga, N. Manini, G.E. Santoro, and M. Sigrist.
We thank particularly L. De Leo for providing us his NRG code. This work
has been partly supported by INFM, MIUR (COFIN 2003 and 2004) and FIRB
RBAU017S8R004.


\begin{thebibliography}{99}
\bibitem{nota} In the presence of particle-hole symmetry, the chemical potential 
is strictly zero whatever is the interaction. On the other hand, 
if Fermi liquid theory holds, then $\mathcal{I}m\, \Sigma(\omega)\sim 
\omega^2$. In this case, it turns out that the value of the DOS at the chemical potential 
should not be affected by $U$ and $J$. Therefore, if one finds a different DOS from the 
{\it bare} one, that necessarily implies a breakdown of Fermi liquid theory.
\end{thebibliography}
\end{document}